# A NEW CLASS OF EXACT SOLUTIONS OF THE SCHRÖDINGER EQUATION

E.E. Perepelkin[a], B.I. Sadovnikov[a], N.G. Inozemtseva[b], A.A. Tarelkin[a]

[a] *Faculty of Physics, Lomonosov Moscow State University, Moscow, 119991 Russia*
E. Perepelkin e-mail: pevgeny@jinr.ru, B. Sadovnikov e-mail: sadovnikov@phys.msu.ru,
A. Tarelkin e-mail: tarelkin.aleksandr@physics.msu.ru
[b] *Dubna State University, Moscow region, Moscow, 141980 Russia*
N. Inozemtseva e-mail: nginozv@mail.ru

**Abstract**
The aim of this paper is to find the exact solutions of the Schrödinger equation. As is known, the Schrödinger equation can be reduced to the continuum equation. In this paper, using the non-linear Legendre transform the equation of continuity is linearized. Particular solutions of such a linear equation are found in the paper and an inverse Legendre transform is considered for them with subsequent construction of solutions of the Schrödinger equation. Examples of the classical and quantum systems are considered.

**Introduction**

As a rule, the Schrödinger equation is used for description of quantum systems [1]. However, the fact that the Schrödinger equation can be reduced to the equation of continuity for the probability density function $f(\vec{r},t)$ indicates the possibility of applying it to the description of the classical systems of continuum mechanics [2-4].

When considering quantum systems, the following consequence of actions is usually performed [5-8]:

- the potential $U(\vec{r},t)$ is specified which enters the Schrödinger equation as well as the initial and boundary conditions for the wave function $\Psi(\vec{r},t)$;

- the Schrödinger equation is solved and a wave function is found, which gives the probability density function $f(\vec{r},t) = |\Psi(\vec{r},t)|^2$;

- knowing the wave function $\Psi = |\Psi|e^{i\varphi}$, a vector field of the probability flux density $\vec{J} = f\langle\vec{v}\rangle = -\frac{i\hbar}{2m}(\bar{\Psi}\nabla\Psi - \Psi\nabla\bar{\Psi})$ is constructed;

- the quantum potential $Q(\vec{r},t) = -\frac{\hbar^2}{2m}\frac{\Delta|\Psi|}{|\Psi|}$ is defined, which is used in the pilot wave theory of de Brogile-Bohm [9-12];

- the «classical» potential $e\chi = U + Q$ is found entering the Hamilton-Jacobi equation $-\hbar\varphi_t = \frac{m|\langle\vec{v}\rangle|^2}{2} + e\chi$.

As a result, we can find the equations of motions for the «classical analog» of the quantum system [9-14].

Papers [3, 4] consider the inverse problem, that is the construction of the Schrödinger equation solution out of the continuity equation solutions. Initially, the vector field of continuous medium velocities $\langle\vec{v}\rangle(\vec{r},t)$ and the probability density function $f_0(\vec{r}) = f(\vec{r},t_0)$ are defined *only* at the initial instant of time $t_0$. According to these data, the potentials $U$ and $Q$, the wave function phase $\varphi$ and the wave function $\Psi$ itself are found. Thus, the «quantum analog» of the classical system is obtained.

In this paper we consider a similar inverse problem, but not the vector field $\langle\vec{v}\rangle$ is defined but the dependence of the probability density function $f$ on the velocity field $|\langle\vec{v}\rangle|$, that is $f = f(|\langle\vec{v}\rangle|)$. The velocity field $\langle\vec{v}\rangle$ itself is considered unknown. The vector field $\langle\vec{v}\rangle$ is supposed to be stationary and two-dimensional, that is $\langle\vec{v}\rangle = \langle\vec{v}\rangle(x,y)$.

The paper has the following structure. In §1, we describe the main method for constructing the solution indicated above for the inverse problem for the dependence $f = f(|\langle\vec{v}\rangle|)$. As an example of the function $f$, we consider a dependence in the form of a Gaussian distribution $\exp(-|\langle\vec{v}\rangle|^2/2\sigma^2)$. As a result, the problem is reduced to a non-linear partial differential equation for the phase of the wave function. Using the non-linear Legendre transform [15-17], we reduce the non-linear equation to a linear partial differential equation. Particular solutions of such an equation can be found by the separation of variables method. After separating the variables two equations are obtained. The first equation has a solution in the form of a system of trigonometric functions.

The solution of the second equation is found in the form of an expansion in a neighborhood of a singular point. In §2, we obtain a general form of the coefficients of such a series and consider the question of convergence of the series.

In §3, we consider questions of summation of the series obtained in §2. It is shown that for a certain choice of parameters finite series corresponding to generalized Laguerre polynomials are obtained.

In §4 we construct the inverse Legendre transform for the solutions obtained in §2-3. It is shown that not all solutions for the inverse Legendre transform give single-valued probability density functions. The function of the wave function phase is a multivalent function in the general case. A single-valued transformation is possible only for certain regions. Expressions for the potentials $U$, $Q$ and for the probability density function are obtained $f$. The vector field map of the velocities $\langle\vec{v}\rangle$ is constructed. We considered limit cases $\sigma \gg \dfrac{\hbar}{2m}$ and $\sigma \sim \dfrac{\hbar}{2m}$ for the classical and quantum systems.

The Conclusion contains the main results of the work.

## §1 Solution construction method

In paper [3], the relation between the continuity equation (the first equation in the Vlasov equation chain) and the Schrödinger equation was considered. As a result, one can construct solutions of the Schrödinger equation from the solutions of the continuity equation. If the probability density function $f(\vec{r},t)$ and the velocity of the probability flow $\langle\vec{v}\rangle(\vec{r},t)$ are known, one can find the quantum potential $Q(\vec{r},t)$ [9-12], the potential $U(\vec{r},t)$ for the Schrödinger equation, the phase $\varphi(\vec{r},t)$ of the wave function, the vector potential $\vec{A}(\vec{r},t)$ and the wave function $\Psi(\vec{r},t) = \sqrt{f(\vec{r},t)}e^{i\varphi(\vec{r},t)}$ itself. The following expressions are true [3]

$$\langle \vec{v} \rangle (\vec{r},t) = i\alpha \nabla \mathrm{Ln}\left[\frac{\Psi}{\bar{\Psi}}\right] + \gamma \vec{A}, \qquad (1.1)$$

where $\nabla \mathrm{Ln}\left[\frac{\Psi}{\bar{\Psi}}\right] = \nabla \Phi(\vec{r},t)$ corresponds to the irrotational one, and $\vec{A}$ - to the solenoidal component of the velocity of the probability flow. The constants $\alpha$ and $\gamma$ are equal to $-\frac{\hbar}{2m}$ and $-\frac{e}{m}$ in the case of considering the quantum system. In this way, the continuum equation for the probability density function $f(\vec{r},t)$:

$$\frac{\partial f(\vec{r},t)}{\partial t} + \mathrm{div}_r \left[ f(\vec{r},t) \langle \vec{v} \rangle (\vec{r},t) \right] = 0, \qquad (1.2)$$

becomes the equation [3, 13-14]:

$$\frac{i}{\beta}\frac{\partial \Psi}{\partial t} = -\alpha\beta \left(\hat{p} - \frac{\gamma}{2\alpha\beta}\vec{A}\right)^2 \Psi + \frac{1}{2\alpha\beta}\frac{|\gamma\vec{A}|^2}{2}\Psi + U\Psi \qquad (1.3)$$

where $\hat{p} = -\frac{i}{\beta}\nabla$, $\beta = \frac{1}{\hbar}$, and the potential $U$ has the following form:

$$U(\vec{r},t) = -\frac{1}{\beta}\left\{\frac{\partial \varphi(\vec{r},t)}{\partial t} + \alpha\left[\frac{\Delta\sqrt{f(\vec{r},t)}}{\sqrt{f(\vec{r},t)}} - |\nabla\varphi(\vec{r},t)|^2\right] + \gamma(\vec{A}, \nabla\varphi)\right\}, \qquad (1.4)$$

where $\varphi(\vec{r},t)$ is the phase of the wave function $\Psi(\vec{r},t)$, which is directly connected with the scalar potential $\Phi(\vec{r},t)$ of the velocity of the probability flow (1.1), as

$$\mathrm{Arg}\left[\frac{\Psi(\vec{r},t)}{\bar{\Psi}(\vec{r},t)}\right] = 2\varphi(\vec{r},t) + 2\pi k = \Phi(\vec{r},t), \qquad (1.5)$$

as

$$\mathrm{Ln}\left[\frac{\Psi}{\bar{\Psi}}\right] = \ln\left|\frac{\Psi}{\bar{\Psi}}\right| + i\,\mathrm{Arg}\left[\frac{\Psi}{\bar{\Psi}}\right] = i\Phi(\vec{r},t). \qquad (1.6)$$

In [3, 13-14] the following equation was obtained (1.4):

$$\frac{\partial \Phi}{\partial t} = -\frac{2}{\hbar}\left[\frac{m|\langle\vec{v}\rangle|^2}{2} + e\chi\right] = -\frac{2}{\hbar}W(\vec{r},t), \qquad (1.7)$$

where $e\chi$ is the potential energy, $T = \frac{m|\langle\vec{v}\rangle|^2}{2}$ is kinetic, and $W(\vec{r},t)$ is total energy of the system. The potential $U$ (1.4) is associated with the classical potential $e\chi$ (1.7) with ratio

$$\chi \stackrel{\text{det}}{=} \frac{2\alpha\beta}{\gamma}\left(\frac{1}{2\alpha\beta}\frac{|\gamma\vec{A}|^2}{2}+U+Q\right), \tag{1.8}$$

where

$$Q = \frac{\alpha}{\beta}\frac{\Delta\sqrt{f}}{\sqrt{f}} = \frac{\alpha}{2\beta}\left(\Delta S + \frac{1}{2}|\nabla S|^2\right), \quad S = \ln f.$$

Note that expression (1.8) contains a summand $Q$ which is a well-known quantum potential in the theory of pilot wave [9-14]. Thus in (1.8) we can find the term $\frac{|\gamma\vec{A}|^2}{2}$ corresponding to the kinetic energy of the solenoidal velocity field. The potentials (1.4) and (1.8) are obtained in [3] based only on equation (1.2) and the representation (1.1).

In this paper we will consider the probability density function of the form

$$f(x,y) = Ce^{-\frac{|\langle\vec{v}\rangle(x,y)|^2}{2\sigma^2}} \stackrel{\text{det}}{=} \mathrm{f}\left(|\langle\vec{v}\rangle(x,y)|\right), \tag{1.9}$$

$$S(x,y) = \ln f(x,y) = \ln \mathrm{f}\left(|\langle\vec{v}\rangle|\right) \stackrel{\text{det}}{=} \mathrm{S}\left(|\langle\vec{v}\rangle|\right) = c - \frac{|\langle\vec{v}\rangle(x,y)|^2}{2\sigma^2},$$

where $c, C, \sigma$ are constant values. Let us assume that there is no vortex flow $\vec{A}$ in the representation (1.1), that is

$$\langle\vec{v}\rangle(x,y) = -\alpha\nabla\Phi(x,y). \tag{1.10}$$

Considering (1.9) and (1.10), the equation (1.2) is of the form

$$\mathrm{div}\left[\mathrm{f}\left(|\alpha\nabla\Phi(x,y)|\right)\nabla\Phi(x,y)\right] = 0,$$

$$\left[|\nabla\Phi| + |\alpha|\Phi_x^2\,\mathrm{S}'(|\alpha\nabla\Phi|)\right]\Phi_{xx} + 2|\alpha|\Phi_x\Phi_y\,\mathrm{S}'(|\alpha\nabla\Phi|)\Phi_{xy} + \left[|\nabla\Phi| + |\alpha|\Phi_y^2\,\mathrm{S}'(|\alpha\nabla\Phi|)\right]\Phi_{yy} = 0,$$

considering that $\mathrm{S}'(|\alpha\nabla\Phi|) = -\frac{|\alpha|}{\sigma^2}|\nabla\Phi|$, we obtain

$$\left[1 - \frac{\alpha^2}{\sigma^2}\Phi_x^2\right]\Phi_{xx} - 2\frac{\alpha^2}{\sigma^2}\Phi_x\Phi_y\Phi_{xy} + \left[1 - \frac{\alpha^2}{\sigma^2}\Phi_y^2\right]\Phi_{yy} = 0, \tag{1.11}$$

Equation (1.11) is a non-linear partial equation with respect to the phase of the wave function (1.5). In the case when

$$\frac{|\alpha|}{\sigma} = \frac{\hbar}{2m\sigma} \ll 1 \tag{1.12}$$

the equation (1.11) can be considered as linear Laplace equation $\Delta\Phi = 0$. In this case the phase $\varphi$ of the wave function $\Psi$ is a harmonic function.

If the condition (1.12) is not satisfied, then it is necessary to solve the non-linear equation (1.11). For the equation (1.11) to be linearized, let us use the Legendre transform [15-17]

$$\omega(\xi,\eta) + \Phi(x,y) = x\xi + y\eta,$$
$$\xi = \Phi_x, \quad \eta = \Phi_y, \quad x = \omega_\xi, \quad y = \omega_\eta, \quad (1.13)$$
$$\Phi_{xx} = \rho\omega_{\eta\eta}, \quad \Phi_{xy} = -\rho\omega_{\xi\eta}, \quad \Phi_{yy} = \rho\omega_{\xi\xi},$$

where $\rho = \Phi_{xx}\Phi_{yy} - \Phi_{xy}^2 = \dfrac{1}{\omega_{\xi\xi}\omega_{\eta\eta} - \omega_{\xi\eta}^2}$ is a Jacobian of the Legendre transform. Stated geometrically, the function $\omega$ represents a set of tangent planes to the surface of the wave function phase. Notice that from the physical standpoint the phase of the wave function can be interpreted as an action.

Applying the Legendre transform (1.13) to the equation (1.11), we obtain a linear equation with respect to the function $\omega$

$$\left(1 - \frac{\alpha^2}{\sigma^2}\xi^2\right)\omega_{\eta\eta} + 2\frac{\alpha^2}{\sigma^2}\xi\eta\omega_{\xi\eta} + \left(1 - \frac{\alpha^2}{\sigma^2}\eta^2\right)\omega_{\xi\xi} = 0. \quad (1.14)$$

In the equation (1.14) it is convenient to substitute the variables

$$\bar{\xi} = \frac{\alpha}{\sigma}\xi, \quad \bar{\eta} = \frac{\alpha}{\sigma}\eta, \quad \bar{\omega}(\bar{\xi},\bar{\eta}) = \omega(\xi,\eta),$$
$$\omega_{\xi\xi} = \frac{\alpha^2}{\sigma^2}\bar{\omega}_{\bar{\xi}\bar{\xi}}, \quad \omega_{\xi\eta} = \frac{\alpha^2}{\sigma^2}\bar{\omega}_{\bar{\xi}\bar{\eta}}, \quad \omega_{\eta\eta} = \frac{\alpha^2}{\sigma^2}\bar{\omega}_{\bar{\eta}\bar{\eta}}. \quad (1.15)$$

With the substitution (1.15) the equation (1.14) takes the form

$$\left(1 - \bar{\xi}^2\right)\bar{\omega}_{\bar{\eta}\bar{\eta}} + 2\bar{\xi}\bar{\eta}\bar{\omega}_{\bar{\xi}\bar{\eta}} + \left(1 - \bar{\eta}^2\right)\bar{\omega}_{\bar{\xi}\bar{\xi}} = 0. \quad (1.16)$$

Let us proceed in the equation (1.16) from Cartesian coordinate system $(\bar{\xi},\bar{\eta})$ to the polar coordinate system $(\tau,\theta)$

$$\bar{\xi} = \tau\cos\theta, \quad \bar{\eta} = \tau\sin\theta, \quad \bar{\omega}(\bar{\xi},\bar{\eta}) \stackrel{\text{det}}{=} \tilde{\omega}(\tau,\theta), \quad (1.17)$$

$$\bar{\omega}_{\bar{\xi}\bar{\xi}} = \tilde{\omega}_{\tau\tau}\cos^2\theta + \tilde{\omega}_{\theta\theta}\frac{\sin^2\theta}{\tau^2} - \tilde{\omega}_{\tau\theta}\frac{\sin 2\theta}{\tau} + \tilde{\omega}_\tau\frac{\sin^2\theta}{\tau} + \tilde{\omega}_\theta\frac{\sin 2\theta}{\tau^2},$$

$$\bar{\omega}_{\bar{\xi}\bar{\eta}} = \tilde{\omega}_{\tau\tau}\frac{\sin 2\theta}{2} - \tilde{\omega}_{\theta\theta}\frac{\sin 2\theta}{2\tau^2} + \tilde{\omega}_{\tau\theta}\frac{\cos 2\theta}{\tau^2} - \tilde{\omega}_\tau\frac{\sin 2\theta}{2\tau} - \tilde{\omega}_\theta\frac{\cos 2\theta}{\tau^2},$$

$$\bar{\omega}_{\bar{\eta}\bar{\eta}} = \tilde{\omega}_{\tau\tau}\sin^2\theta + \tilde{\omega}_{\theta\theta}\frac{\cos^2\theta}{\tau^2} + \tilde{\omega}_{\tau\theta}\frac{\sin 2\theta}{\tau} + \tilde{\omega}_\tau\frac{\cos^2\theta}{\tau} - \tilde{\omega}_\theta\frac{\sin 2\theta}{\tau^2}.$$

Substituting (1.17) into the equation (1.16), we obtain

$$\tilde{\omega}_{\tau\tau} + a(\tau)\left[\frac{1}{\tau}\tilde{\omega}_\tau + \frac{1}{\tau^2}\tilde{\omega}_{\theta\theta}\right] = 0, \qquad (1.18)$$

$$a(\tau) \stackrel{det}{=} 1 - \tau^2.$$

Let us find the particular solutions of the equation (1.18) by the separation of variables method. Substituting $\tilde{\omega} \sim T(\tau)\Theta(\theta)$ into the equation (1.18), we obtain

$$T'' + a(\tau)\left[\frac{1}{\tau}T' - \frac{\lambda^2}{\tau^2}T\right] = 0, \qquad (1.19)$$

$$\Theta'' + \lambda^2\Theta = 0, \qquad (1.20)$$

where $\lambda$ is a constant value. The solution of the equation (1.20) for the function $\Theta(\theta)$ can be represented as the system of the functions $\sin\lambda\theta$ and $\cos\lambda\theta$. The solution of the equation (1.19) for the function $T(\tau)$ can be represented as the series expansion in the neighborhood of the singular point $\tau_0 = 0$.

§2 **Representation of the solution in the form of a series**
We will solve the equation (1.19) in the form of an expansion in the neighborhood of the singular point $\tau_0 = 0$

$$T(\tau) = \tau^\nu \sum_{k=0}^{+\infty} a_k \tau^k. \qquad (2.1)$$

Substituting (2.1) into the equation (1.19) and equating the coefficients of the same powers $\tau$, we obtain

$$T'(\tau) = \sum_k a_k(k+\nu)\tau^{k+\nu-1}, \quad T''(\tau) = \sum_k a_k(k+\nu)(k+\nu-1)\tau^{k+\nu-2},$$

$$\sum_k a_k(k+\nu)(k+\nu-1)\tau^{k+\nu-2} + \left(\frac{1}{\tau} - \tau\right)\sum_k a_k(k+\nu)\tau^{k+\nu-1} - \left(\frac{\lambda^2}{\tau^2} - \lambda^2\right)\sum_k a_k\tau^{k+\nu} = 0,$$

$$\sum_k a_k(k+\nu)(k+\nu-1)\tau^{k+\nu-2} + \sum_k a_k(k+\nu)\tau^{k+\nu-2} - \sum_k a_k(k+\nu)\tau^{k+\nu} -$$

$$-\sum_k \lambda^2 a_k \tau^{k+\nu-2} + \sum_k a_k \lambda^2 \tau^{k+\nu} = 0,$$

$$\sum_k a_k\left[(k+\nu)(k+\nu-1) + (k+\nu) - \lambda^2\right]\tau^{k+\nu-2} - \sum_k a_k\left[(k+\nu) - \lambda^2\right]\tau^{k+\nu} = 0,$$

$$\sum_k a_k\left[(k+\nu)^2 - \lambda^2\right]\tau^{k+\nu-2} - \sum_k a_k\left[(k+\nu) - \lambda^2\right]\tau^{k+\nu} = 0. \qquad (2.2)$$

Equating the coefficients for $\tau^{\nu-2}$ gives the expression for $\nu$

$$\nu = \pm|\lambda|, \qquad (2.3)$$

provided that $a_0 \neq 0$. With the power $\tau^{\nu-1}$, we obtain

$$a_1\left[(1+\nu)^2-\nu^2\right]=0 \Rightarrow \begin{cases} a_1=0, \\ \nu=-\dfrac{1}{2}. \end{cases} \qquad (2.4)$$

In the case $\nu=-\dfrac{1}{2}$, the singular solutions (1.19) will be obtained at the point $\tau=0$, therefore, in what follows we consider the case $a_1=0$.

For an arbitrary coefficient $a_k$ with an index $k\geq 2$ from the expression (2.2) one can obtain a recurrence relation. By changing the variable $s=k-2$, we obtain

$$\sum_{s=0}^{+\infty} a_{s+2}\left[(s+2+\nu)^2-\lambda^2\right]\tau^{s+\nu}-\sum_{k=0}^{+\infty} a_k\left[(k+\nu)-\lambda^2\right]\tau^{k+\nu}=0,$$

or

$$a_{k+2}\left[(k+2+\nu)^2-\lambda^2\right]=a_k\left[(k+\nu)-\lambda^2\right],$$

$$a_{k+2}=a_k\frac{k+\nu-\nu^2}{(k+2+\nu)^2-\nu^2}=a_k\frac{k+\nu-\nu^2}{(k+2)^2+2\nu(k+2)},$$

$$a_{k+2}=a_k\frac{k+\nu-\nu^2}{(k+2)(k+2(\nu+1))},\quad k\geq 2. \qquad (2.5)$$

As the coefficient $a_1=0$, then from (2.5) it follows that the odd coefficients equal to zero

$$a_1=a_3=a_5=\ldots=a_{2s+1}=0,\quad s=0,1,2\ldots \qquad (2.6)$$

From (2.6) it follows that in the series (2.1) only even coefficients $a_{2s}$, $s=0,1,2\ldots$, as $a_0\neq 0$ (2.3) remain. As a result, the expression (2.5) for the even coefficients takes the form

$$a_{2s+2}=a_{2s}\frac{2s+\nu-\nu^2}{4(s+1)(s+\nu+1)},\quad s=0,1,2,\ldots \qquad (2.7)$$

or introducing a new designation

$$b_0=const\neq 0,$$
$$b_{s+1}=b_s\frac{2s+\nu-\nu^2}{4(s+1)(s+\nu+1)},\quad s=0,1,2,\ldots \qquad (2.8)$$

Considering (2.8), the expression (2.1) takes the form

$$T(\tau)=\tau^\nu\sum_{s=0}^{+\infty}b_s\tau^{2s}=\tau^\nu\sum_{s=0}^{+\infty}b_s\tau^{2s}. \qquad (2.9)$$

From (2.8) we obtain the recurrence relation for the coefficient $b_s$ in terms of the coefficient $b_0$

$$b_1 = b_0 \frac{v - v^2}{4(1+v)},$$

$$b_2 = b_1 \frac{2+v-v^2}{2 \cdot 4(2+v)} = b_0 \frac{(v-v^2)(2+v-v^2)}{4^2 2(1+v)(2+v)},$$

$$b_3 = b_2 \frac{4+v-v^2}{4 \cdot 3(3+v)} = b_0 \frac{(v-v^2)(2+v-v^2)(4+v-v^2)}{4^3(1+v)(2+v)(3+v) 2 \cdot 3},$$

$$b_4 = b_3 \frac{6+v-v^2}{4 \cdot 4(4+v)} = b_0 \frac{(v-v^2)(2+v-v^2)(4+v-v^2)(6+v-v^2)}{4^4(1+v)(2+v)(3+v)(4+v) 2 \cdot 3 \cdot 4},$$

...

$$b_s = \frac{b_0 v}{4^s (v)_{s+1} s!} \prod_{k=0}^{s-1}(2k+v-v^2), \; s \geq 1 \qquad (2.10)$$

$$b_s = \frac{b_0}{4^s s!} \prod_{k=0}^{s-1} \frac{2k+v-v^2}{v+k+1}, \; s \geq 1$$

The variable $b_0$ in the expression (2.10) can be taken equal to unity or chosen, for example, from the normalization condition. Thus, the solution of the equation (1.19) can be presented as a superposition of solutions of the type (2.9)

$$T^{(1,2)}(\tau) = \tau^{v_{1,2}} b_0 + \tau^{v_{1,2}} \sum_{s=1}^{+\infty} \frac{b_0}{s!} \left(\frac{\tau}{2}\right)^{2s} \prod_{k=0}^{s-1} \frac{2k+v_{1,2}-v_{1,2}^2}{v_{1,2}+k+1}, \qquad (2.11)$$

where $v_{1,2} = \pm |\lambda|$.

Let us find the convergence radius of the series (2.11) for a fixed $v$. According to the d'Alembert's ratio test, we have

$$q = \lim_{s \to +\infty} \left| \frac{b_{s+1} \tau^{2s+2}}{b_s \tau^{2s}} \right| = \frac{\tau^2}{2} \lim_{s \to +\infty} \left| \frac{2 + \frac{v-v^2}{s}}{s \left(1 + \frac{1}{s}\right)\left(1 + \frac{v+1}{s}\right)} \right| = 0 < 1.$$

Thus, the convergence radius of the series (2.11) is of the form

$$0 \leq \tau < +\infty. \qquad (2.12)$$

### §3. Particular solutions

Let us consider particular solutions of the equation (1.18). The function $\Theta(\theta)$ satisfies an equation

$$\Theta'' + \lambda^2 \Theta = 0. \qquad (3.1)$$

In the degenerate case with $\lambda = 0$, the solutions (3.1) and (2.11) are of the form

$$\Theta(\theta) = c_1\theta + c_2, \qquad (3.2)$$
$$T(\tau) = b_0,$$

where $c_1, c_2$ are constant values. Consequently, the function $\tilde{\omega}(\tau, \theta)$ takes the form

$$\tilde{\omega}(\tau, \theta) = C_1\theta + C_2. \qquad (3.3)$$

With, $\lambda \neq 0$ the solution (3.1) can be represented as a superposition

$$\Theta(\theta) = \begin{cases} \sin \lambda\theta \\ \cos \lambda\theta \end{cases}, \qquad (3.4)$$

where $\lambda$ is defined from the boundary conditions. Let us consider different types of the boundary conditions.

### 3.1 Periodic boundary condition

Let the following condition is satisfied

$$\Theta(0) = \Theta(2\pi), \qquad (3.5)$$

then

$$\lambda_n = n, \ n \in \mathbb{N}. \qquad (3.6)$$

Considering the bounded solutions ($v_1 = |\lambda|$) from (3.6) and (2.11), we obtain

$$T_n(\tau) = b_0\tau^n + \tau^n \sum_{s=1}^{+\infty} \frac{b_0}{s!} \left(\frac{\tau}{2}\right)^{2s} \prod_{k=0}^{s-1} \frac{2k+n-n^2}{n+k+1}. \qquad (3.7)$$
$$\tilde{\omega}_n(\tau, \theta) = T_n(\tau)(A_n \sin n\theta + B_n \cos n\theta).$$

For $n = 1$ the expression (3.7) is of the form

$$T_1(\tau) = b_0\tau. \qquad (3.8)$$

**Remark**

Consequently, the function $\tilde{\omega}_1(\tau, \theta)$ corresponds to the equation of a plane

$$\tilde{\omega}_1(\tau, \theta) = A\tau \sin\theta + B\tau \cos\theta = A\bar{\eta} + B\bar{\xi} = \frac{\alpha}{\sigma}(A\eta + B\xi), \qquad (3.9)$$

For which the Legendre transform is impossible, as the Jacobian of the transform (1.13) is unbounded.

For $n = 2$, the expression (3.7) takes the form

$$T_2(\tau) = b_0\tau^2 + \tau^2 \sum_{s=1}^{+\infty} \frac{b_0}{s!}\left(\frac{\tau}{2}\right)^{2s} \prod_{k=0}^{s-1} \frac{2k-2}{k+3},$$

$$T_2(\tau) = b_0\tau^2 - \frac{b_0}{6}\tau^4 = b_0\tau^2\left(1 - \frac{\tau^2}{6}\right). \tag{3.10}$$

The function $T_2(\tau)$ has two zeros for $\tau = 0$ and $\tau = \sqrt{6}$.

For $n = 3$

$$T_3(\tau) = b_0\tau^3 + \tau^3\sum_{s=1}^{+\infty}\frac{b_0}{s!}\left(\frac{\tau}{2}\right)^{2s}\prod_{k=0}^{s-1}\frac{2k-6}{k+4} = b_0\tau^3 - \tau^3\frac{3b_0}{2}\left(\frac{\tau}{2}\right)^2 + \tau^3\frac{3b_0}{5}\left(\frac{\tau}{2}\right)^4 - \tau^3\frac{b_0}{15}\left(\frac{\tau}{2}\right)^6,$$

$$T_3(\tau) = b_0\tau^3\left(1 - \frac{3}{8}\tau^2 + \frac{3}{80}\tau^4 - \frac{1}{960}\tau^6\right). \tag{3.11}$$

The following solutions can be obtained in a similar way $T_n(\tau)$.

There is another way to find the solutions of the equation (1.19) in the case (3.6). The solution (2.11) can be represented as generalized Laguerre polynomials [18, 19]. Let us find the solution of the equation (1.19) in the form

$$T(\tau) = \tau^\lambda G(t) = \tau^\lambda G\left(\frac{\tau^2}{2}\right). \tag{3.12}$$

Let us use (3.12) in (1.19), we obtain

$$T'(\tau) = \lambda\tau^{\lambda-1}G(t) + G'(t)\tau^{\lambda+1} = \lambda\left(\sqrt{2t}\right)^{\lambda-1}G(t) + \left(\sqrt{2t}\right)^{\lambda+1}G'(t),$$

$$T''(\tau) = \lambda(\lambda-1)\tau^{\lambda-2}G(t) + \lambda\tau^\lambda G'(t) + (\lambda+1)G'(t)\tau^\lambda + G''(t)\tau^{\lambda+2} =$$
$$= \lambda(\lambda-1)\tau^{\lambda-2}G(t) + (2\lambda+1)G'(t)\tau^\lambda + G''(t)\tau^{\lambda+2} =$$
$$= \lambda(\lambda-1)\left(\sqrt{2t}\right)^{\lambda-2}G(t) + (2\lambda+1)G'(t)\left(\sqrt{2t}\right)^\lambda + G''(t)\left(\sqrt{2t}\right)^{\lambda+2},$$

$$T'' + a(\tau)\left[\frac{1}{\tau}T' - \frac{\lambda^2}{\tau^2}T\right] = \lambda(\lambda-1)\left(\sqrt{2t}\right)^{\lambda-2}G + (2\lambda+1)G'\left(\sqrt{2t}\right)^\lambda + G''\left(\sqrt{2t}\right)^{\lambda+2} +$$

$$+ (1-2t)\left(\frac{1}{\sqrt{2t}}\left(\lambda\left(\sqrt{2t}\right)^{\lambda-1}G + \left(\sqrt{2t}\right)^{\lambda+1}G'\right) - \frac{\lambda^2}{2t}\left(\sqrt{2t}\right)^\lambda G\right) =$$

$$\lambda(\lambda-1)G + (2\lambda+1)G'\left(\sqrt{2t}\right)^2 + G''\left(\sqrt{2t}\right)^4 + (1-2t)\left(\left(\sqrt{2t}\right)^2 G' + \lambda G - \lambda^2 G\right) = 0,$$

$$G''4t^2 + 2t(2\lambda+1)G' + (1-2t)2tG' + \lambda(\lambda-1)G - (1-2t)\lambda(\lambda-1)G = 0,$$

$$G''4t^2 + 4tG'[1+\lambda-t] + 2t\lambda(\lambda-1)G = 0,$$

$$tG'' + (1+\lambda-t)G' + \frac{\lambda(\lambda-1)}{2}G = 0. \tag{3.13}$$

Under the condition (3.6) the solution of the equation (3.13) is generalized Laguerre polynomials

$$G(t) = L_{\frac{n(n-1)}{2}}^{(n)}(t). \qquad (3.14)$$

As a result, the solution of the equation (1.19) under the condition (3.6) takes the form

$$T_n(\tau) = \tau^n L_{\frac{n(n-1)}{2}}^{(n)}\left(\frac{\tau^2}{2}\right). \qquad (3.15)$$

The obtained solution (3.15) coincides with the solution (2.11) under the condition (3.6). Indeed,

$$\begin{aligned}
&n=1: T_1(\tau) = \tau L_0^{(1)}\left(\frac{\tau^2}{2}\right) = \tau, \\
&n=2: T_2(\tau) = \tau^2 L_1^{(2)}\left(\frac{\tau^2}{2}\right) = 3\tau^2\left(1 - \frac{\tau^2}{6}\right), \\
&n=3: T_3(\tau) = \tau^3 L_3^{(3)}\left(\frac{\tau^2}{2}\right) = 20\tau^3\left(1 - \frac{3}{8}\tau^2 + \frac{3}{80}\tau^4 - \frac{\tau^6}{960}\right),
\end{aligned} \qquad (3.16)$$

...

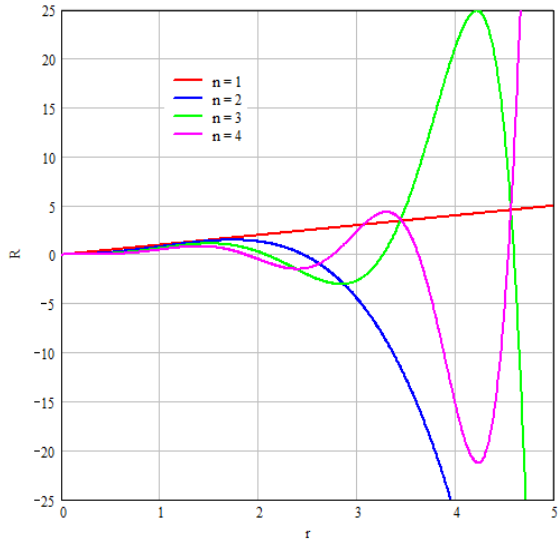

Fig. 1 Graphs of the functions $T_n(\tau)$

As the solution $T_n(\tau)$ is determined up to a constant factor, for example, $c_n = \frac{1}{b_0} L_{\frac{n(n-1)}{2}}^{(n)}(0)$, then the expression (3.16) corresponds to the obtained expressions (3.8), (3.10), (3.11).

Fig. 1 shows the graphs of the functions (3.16) normalized to a value $c_n$ for different values of $n$ and $b_0 = 1$.

Thus, the solution $\tilde{\omega}_n(\tau, \theta)$ is of the form

$$\tilde{\omega}_n(\tau, \theta) = \tau^n L_{\frac{n(n-1)}{2}}^{(n)}\left(\frac{\tau^2}{2}\right)(A_n \sin n\theta + B_n \cos n\theta). \qquad (3.17)$$

### 3.2 Non-periodic boundary conditions

Let us consider the case when $\lambda$ is not an integer. Let us suppose the following condition is satisfied

$$\lambda_j^{(\pm)} = \frac{1 \pm \sqrt{1+8j}}{2}, \quad j = 0, 1, 2, \ldots \qquad (3.18)$$

Values $\lambda_j$ are the roots of the equation $2j - \lambda_j^2 + \lambda_j = 0$. Consequently, the product in (2.11) satisfies the condition

$$\prod_{k=0}^{s-1} \frac{2k + \lambda_j - \lambda_j^2}{\lambda_j + k + 1} = 0 \ \ npu \ \ s \geq j+1. \tag{3.19}$$

If the condition (3.18) is satisfied, (3.19) the number of terms in the series (2.11) is finite and the value $\dfrac{\lambda_j(\lambda_j - 1)}{2}$ is an integer one. Thus, the equation (1.19) has the solutions in the form of generalized Laguerre polynomials, that is

$$T_j(\tau) = \tau^{\lambda_j} L_j^{(\lambda_j)}\left(\frac{\tau^2}{2}\right). \tag{3.20}$$

For values $j = 0,1,3,6,10,...$ the quantities of $\lambda_j$ are integer and the solutions of (3.20) coincide with the corresponding solutions of (3.15). And for values $j = 2,4,5,7,8,9,...$ the quantities of $\lambda_j$ are not integer and new solutions are obtained expressed in terms of generalized Laguerre polynomials (3.20).

Fig.2 shows the solutions of (3.20) corresponding to $\lambda_j^{(+)}$, $j = 2,4,5$, and Fig. 3 shows the solutions of (3.20) corresponding to $\lambda_j^{(-)}$, $j = 2,4,5$.

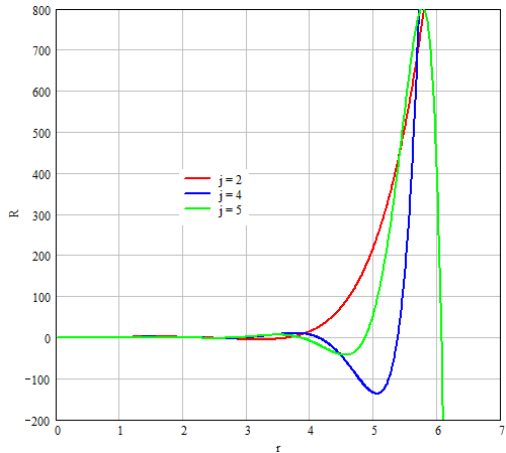

Fig. 2 Graphs of the functions $T_j^{(+)}(\tau)$

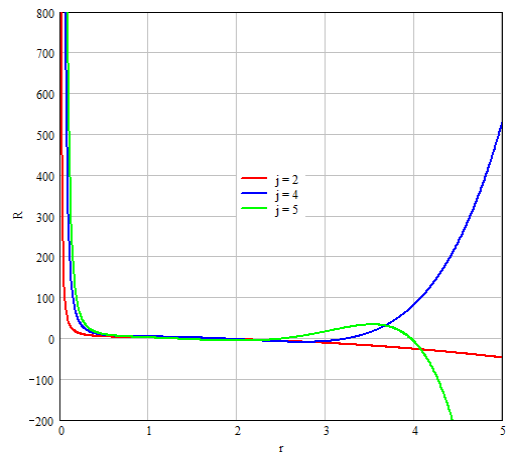

Fig. 3 Graphs of the functions $T_j^{(-)}(\tau)$

If $\lambda$ does not satisfy the condition (3.18), then the number of terms in the series (2.11) is infinite. According to (2.12), the series (2.11) is converges absolutely in the whole space and uniformly on any closed disk.

Figs. 4, 5 show the graphs of the solutions $T_\lambda(\tau)$ (2.11) for the cases $\lambda = \frac{1}{2}, \frac{3}{2}, \frac{5}{2}, \frac{7}{2}$ and $\lambda = -\frac{1}{2}, -\frac{3}{2}, -\frac{5}{2}, -\frac{7}{2}$, respectively.

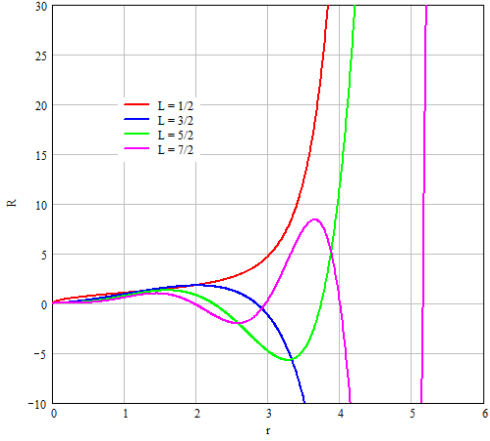 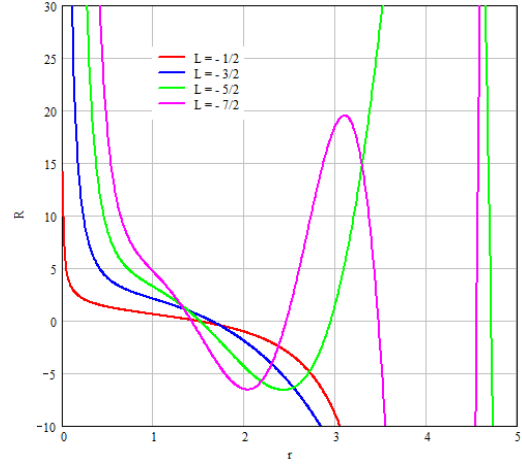

| Fig. 4 Graphs of the functions $T_\lambda(\tau)$ for $\lambda = \frac{1}{2}, \frac{3}{2}, \frac{5}{2}, \frac{7}{2}$ | Fig. 5 Graphs of the functions $T_\lambda(\tau)$ for $\lambda = -\frac{1}{2}, -\frac{3}{2}, -\frac{5}{2}, -\frac{7}{2}$ |
|---|---|

## §4 Inverse Legendre transform

The solutions of $\omega(\xi, \eta)$ obtained in §2-3 must be mapped back according to the Legendre transform (1.13) and the solutions of $\Phi(x, y)$ must be obtained corresponding to the phase of the wave function (1.5). If the functions $\Phi(x, y)$ is known, one can determine the velocity field $\langle \vec{v} \rangle(x, y)$ from the formula (1.1). From the expression (1.9), knowing $\langle \vec{v} \rangle(x, y)$, one can find the probability density function $f(x, y)$. Substituting the probability density $f(x, y)$ into the expressions (1.8) and (1.4) we find the quantum potential $Q(x, y)$ and the potential $U(x, y)$ entering the Schrödinger equation (1.3).

As a result, all necessary quantities from §1 will be determined.

### 4.1 Transformation of the phase and the probability density

To perform the inverse Legendre transform for the phase of the wave function, it is necessary to find the derivatives of the solution $\omega$. Without loss of generality, we consider the solutions $\tilde{\omega}_n(\tau, \theta)$ (3.17)

$$\nabla \tilde{\omega}_n(\tau, \theta) = \Theta_n(\theta) T'_n(\tau) \vec{e}_\tau + \Theta'_n(\theta) \frac{T_n(\tau)}{\tau} \vec{e}_\theta. \tag{4.1}$$

Let us calculate $T'_n(\tau)$, taking into consideration the relation for the derivatives of the generalized Laguerre polynomials [18,19]

$$\frac{d^k}{dt^k} L_n^{(\vartheta)}(t) = (-1)^k L_{n-k}^{(\vartheta+k)}(t), \quad k \leq n,$$

$$T'_n(\tau) = n\tau^{n-1} L_{\frac{n(n-1)}{2}}^{(n)}\left(\frac{\tau^2}{2}\right) - \tau^{n+1} L_{\frac{(n-2)(n+1)}{2}}^{(n+1)}\left(\frac{\tau^2}{2}\right). \tag{4.2}$$

Substituting (4.2) into (4.1), we obtain

$$\nabla \tilde{\omega}_n(\tau,\theta) = \left( n\tau^{n-1} L^{(n)}_{\frac{n(n-1)}{2}}\left(\frac{\tau^2}{2}\right) - \tau^{n+1} L^{(n+1)}_{\frac{(n-2)(n+1)}{2}}\left(\frac{\tau^2}{2}\right) \right)(A_n \sin n\theta + B_n \cos n\theta)\vec{e}_\tau + $$

$$+ \left( \tau^n L^{(n)}_{\frac{n(n-1)}{2}}\left(\frac{\tau^2}{2}\right) \right)(A_n \cos n\theta - B_n \sin n\theta)\vec{e}_\theta. \qquad (4.3)$$

According to the Legendre transform (1.13) and the expressions (1.15), (1.17), the function $\Phi$ can be represented in the form

$$\xi = \frac{\sigma}{\alpha}\tau\cos\theta, \quad \eta = \frac{\sigma}{\alpha}\tau\sin\theta, \qquad (4.4)$$

$$\tau_\xi = \frac{\alpha}{\sigma}\cos\theta, \quad \tau_\eta = \frac{\alpha}{\sigma}\sin\theta, \quad \theta_\xi = -\frac{\alpha}{\sigma}\frac{\sin\theta}{\tau}, \quad \theta_\eta = \frac{\alpha}{\sigma}\frac{\cos\theta}{\tau},$$

$$\frac{\sigma}{\alpha}x = \frac{\sigma}{\alpha}\omega_\xi = \tilde{\omega}_\tau\cos\theta - \tilde{\omega}_\theta\frac{\sin\theta}{\tau}, \quad \frac{\sigma}{\alpha}y = \frac{\sigma}{\alpha}\omega_\eta = \tilde{\omega}_\tau\sin\theta + \tilde{\omega}_\theta\frac{\cos\theta}{\tau},$$

as a result,

$$\Phi(x,y) = -\tilde{\omega}(\tau,\theta) + x\xi + y\eta = $$
$$= -\tilde{\omega}(\tau,\theta) + \left(\tilde{\omega}_\tau\cos\theta - \tilde{\omega}_\theta\frac{\sin\theta}{\tau}\right)\tau\cos\theta + \left(\tilde{\omega}_\tau\sin\theta + \tilde{\omega}_\theta\frac{\cos\theta}{\tau}\right)\tau\sin\theta = $$
$$= -\tilde{\omega} + \tau\tilde{\omega}_\tau\cos^2\theta - \tilde{\omega}_\theta\sin\theta\cos\theta + \tau\tilde{\omega}_\tau\sin^2\theta + \tilde{\omega}_\theta\cos\theta\sin\theta,$$

$$\Phi = -\tilde{\omega} + \tau\tilde{\omega}_\tau = -T(\tau)\Theta(\theta) + \tau T'(\tau)\Theta(\theta) = \Theta(\theta)\left(\tau T'(\tau) - T(\tau)\right). \qquad (4.5)$$

Substituting the expression (3.15) and (4.2) into (4.5), we obtain

$$\tau T'_n(\tau) - T_n(\tau) = n\tau^n L^{(n)}_{\frac{n(n-1)}{2}}\left(\frac{\tau^2}{2}\right) - \tau^{n+2} L^{(n+1)}_{\frac{(n-2)(n+1)}{2}}\left(\frac{\tau^2}{2}\right) - \tau^n L^{(n)}_{\frac{n(n-1)}{2}}\left(\frac{\tau^2}{2}\right) = $$

$$= (n-1)\tau^n L^{(n)}_{\frac{n(n-1)}{2}}\left(\frac{\tau^2}{2}\right) - \tau^{n+2} L^{(n+1)}_{\frac{(n-2)(n+1)}{2}}\left(\frac{\tau^2}{2}\right),$$

$$\Phi_n = \tau^n \left[ (n-1) L^{(n)}_{\frac{n(n-1)}{2}}\left(\frac{\tau^2}{2}\right) - \tau^2 L^{(n+1)}_{\frac{(n-2)(n+1)}{2}}\left(\frac{\tau^2}{2}\right) \right](A_n \sin n\theta + B_n \cos n\theta). \qquad (4.6)$$

According to the Legendre transform and the expression (1.9) and (4.2)

$$|\langle\vec{v}\rangle|^2 = |\alpha\nabla\Phi|^2 = \alpha^2\left(\xi^2 + \eta^2\right) = \sigma^2\tau^2,$$

$$f_n(x,y) = f\left(x_n(\tau,\theta), y_n(\tau,\theta)\right) = Ce^{-\frac{\tau^2}{2}}, \qquad (4.7)$$

where, according to (4.2),

$$x_n(\tau,\theta) = \frac{\alpha}{\sigma}\left[T'_n(\tau)\Theta_n(\theta)\cos\theta - T_n(\tau)\Theta'_n(\theta)\frac{\sin\theta}{\tau}\right],$$
$$y_n(\tau,\theta) = \frac{\alpha}{\sigma}\left[T'_n(\tau)\Theta_n(\theta)\sin\theta + T_n(\tau)\Theta'_n(\theta)\frac{\cos\theta}{\tau}\right]. \tag{4.8}$$

Plotting the graph of the probability density $f_n(x,y)$ for a fixed $n$ is performed according to the following algorithm:
- we take a point $(\tau,\theta)$ and find for it the value of the coordinates $(x,y)$ using the formulas (4.8) and the value of the function $f$ using the formula (4.7);
- we take the next point $(\tau,\theta)$.

For $n=1$ (3.8) the Legendre transform is impossible (see Remark), so we start with the case $n=2$. As the region of the mapping, let us take a disk of a radius $\tau_0 = 2$, $\sigma = 1, b_0 = 1, A_n = 1, B_n = 0$, that is $\tilde{\Omega}_{\tau_0, 2\pi}$, where

$$\tilde{\Omega}_{\tau_0,\theta_0} = \{(\tau,\theta): 0<\tau<\tau_0, 0\leq\theta<\theta_0\}. \tag{4.9}$$

Figs.6-8 show the graphs of the functions $\Phi_2(x,y)$, $|\nabla\Phi_2(x,y)|$ and $f_2(x,y)$, respectively.

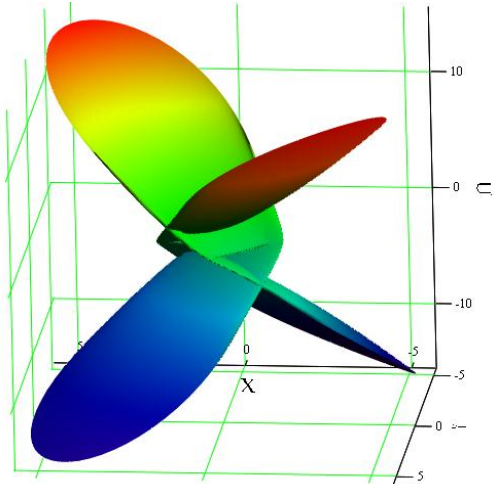

Fig. 6 Graphs of the function $\Phi_2(x,y)$ for $\tilde{\Omega}_{2,2\pi}$

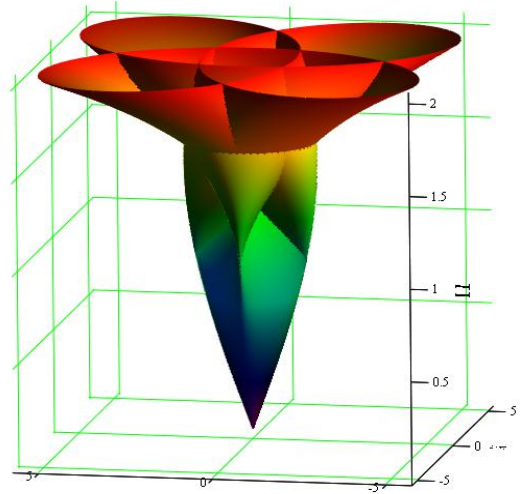

Fig. 7 Graphs of the functions $|\nabla\Phi_2(x,y)|$ for $\tilde{\Omega}_{2,2\pi}$

Figs. 6-8 show that the functions are multivalued when the domain $\tilde{\Omega}_{2,2\pi}$ is transformed. If the radius of the disk is reduced to $\tau_0 = 0.8$, then we obtain single-valued functions (see Figs. 9-11).

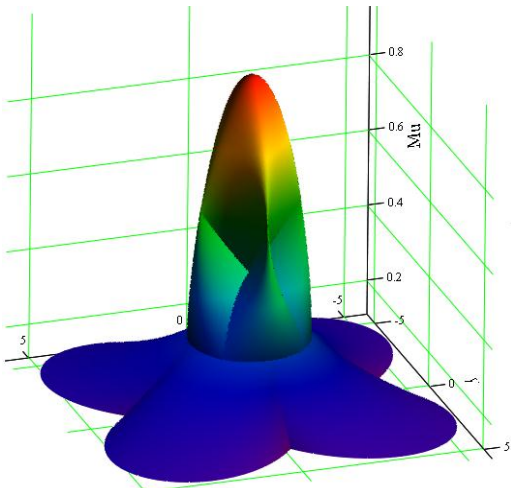
Fig. 8 Graphs of the functions $f_2(x,y)$ for $|\nabla \Phi_2(x,y)|$

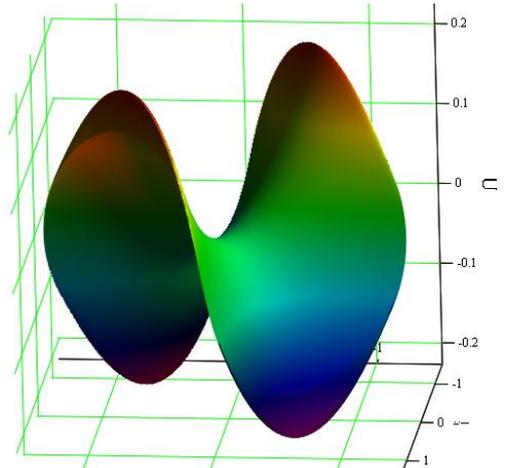
Fig. 9 Graphs of the functions $\Phi_2(x,y)$ for $\tilde{\Omega}_{0.8,2\pi}$

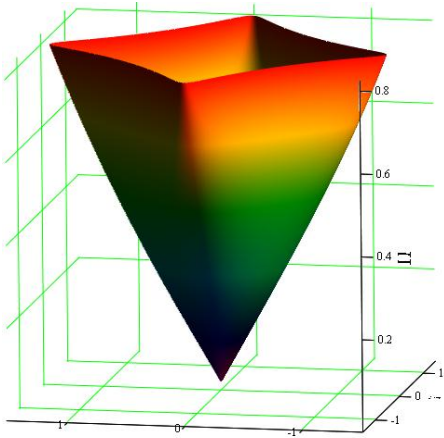
Fig. 10 Graphs of the functions $|\nabla \Phi_2(x,y)|$ for $\tilde{\Omega}_{0.8,2\pi}$

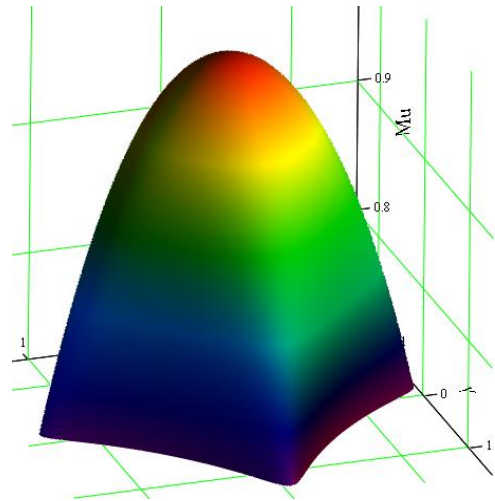
Fig. 11 Graphs of the functions $f_2(x,y)$ for $\tilde{\Omega}_{0.8,2\pi}$

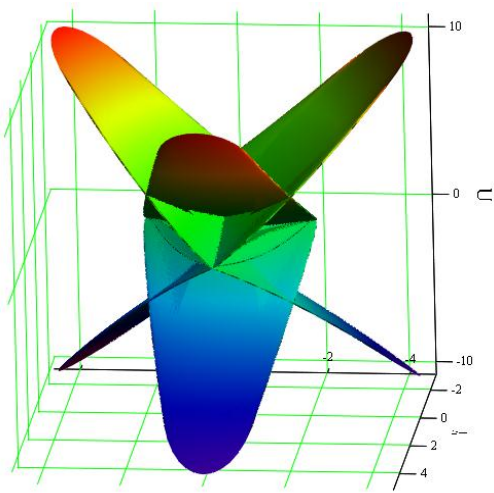
Fig. 12 Graphs of the functions $\Phi_3(x,y)$ for $\tilde{\Omega}_{2,2\pi}$

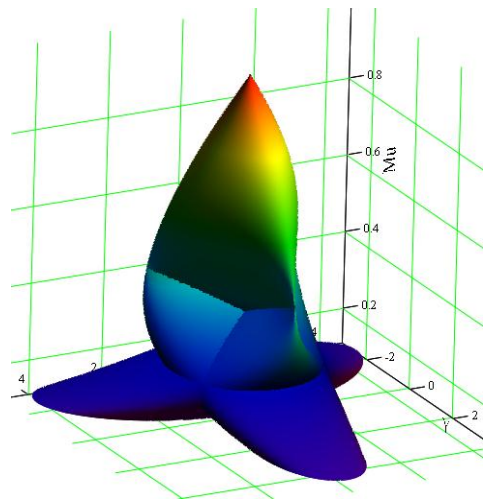
Fig. 13 Graphs of the functions $f_3(x,y)$ for $\tilde{\Omega}_{2,2\pi}$

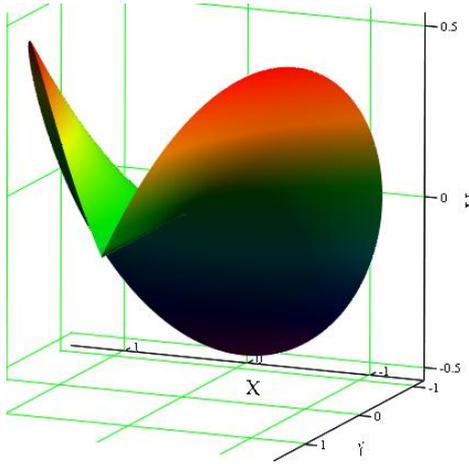

Fig. 14 Graphs of the functions $\Phi_3(x,y)$ for

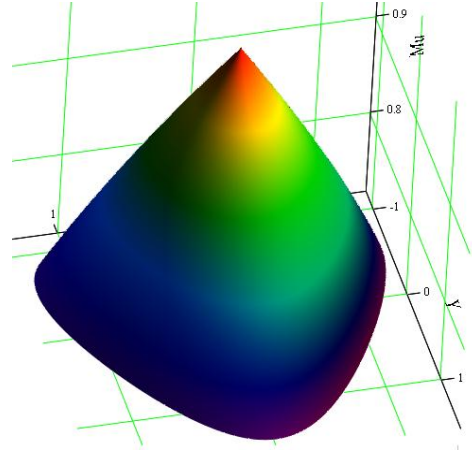

Fig. 15 Graphs of the functions $f_3(x,y)$ for $\tilde{\Omega}_{0.8,\pi}$

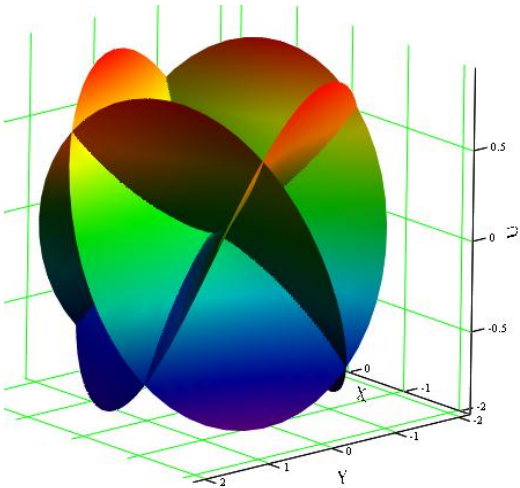

Fig. 16 Graphs of the functions $\Phi_4(x,y)$ for $\tilde{\Omega}_{1,2\pi}$

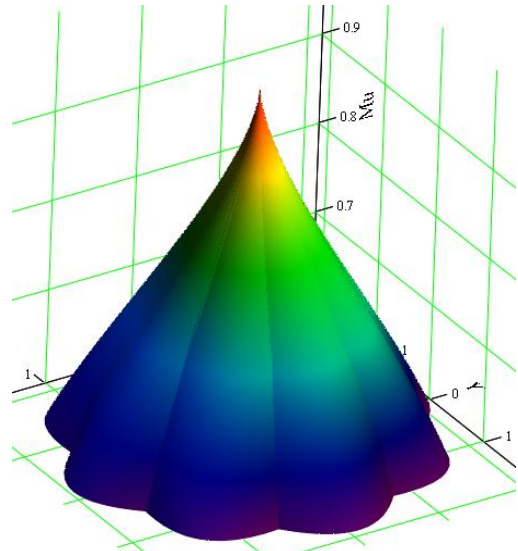

Fig. 17 Graphs of the functions $f_4(x,y)$ for $\tilde{\Omega}_{1,2\pi}$

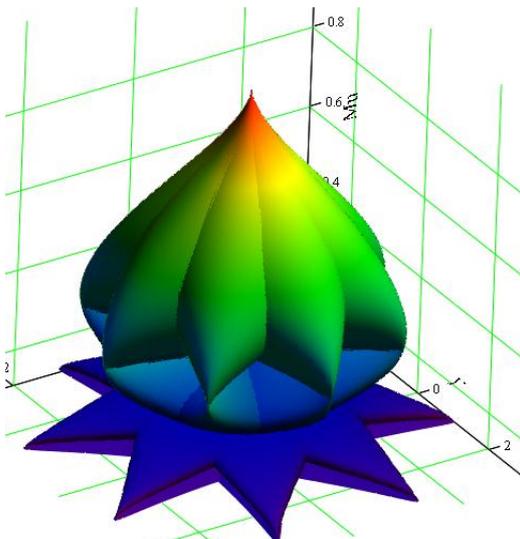

Fig. 18 Graphs of the functions $\Phi_4(x,y)$ for $\tilde{\Omega}_{2,2\pi}$

In the case $n=3$, the domain transformation of the disk $\tilde{\Omega}_{\tau_0,2\pi}$ gives a multivalued function, too (see Figs. 12-13). To the mapping single-valued, it is necessary to map not the domain of the disk but the domain of the sector $\tilde{\Omega}_{\tau_0,\pi}$ with an angle $\theta_0 = \pi$ (see Figs. 14-15).

Fig. 14 shows that the function $\Phi_3(x,y)$ is not smooth in the case $\theta = 0$, but is continuous and single-valued. For subsequent values $n \geq 4$ the function $\Phi_n(x,y)$ can have two main types. The first type when $\Phi_n(x,y)$ is single-valued and has a discontinuity at $\theta = 0$. The second type when $\Phi_n(x,y)$ is multivalued and at the same time is continuous. Figs. 16-17 show the graphs of $\Phi_4(x,y)$ and $f_4(x,y)$ corresponding to the region $\tilde{\Omega}_{0.8,\pi}$, and Fig. 18 shows the graph $f_4(x,y)$, corresponding to the region $\tilde{\Omega}_{0.8,\pi}$.

### 4.2 The transform of the quantum potential

Let us obtain an expression for the inverse transform of the quantum potential Q (1.8), (4.7)

$$Q = \frac{\alpha}{2\beta}\left(\Delta_r S + \frac{1}{2}|\nabla_r S|^2\right) = -\frac{\alpha}{4\beta}\left(\Delta_r \tau^2 - \frac{1}{4}|\nabla_r \tau^2|^2\right). \tag{4.10}$$

Let us calculate $\nabla_r \tau^2$

$$\frac{\sigma^2}{\alpha^2}\nabla_r \tau^2 = 2(\xi\xi_x + \eta\eta_x)\vec{e}_x + 2(\xi\xi_y + \eta\eta_y)\vec{e}_y = 2(\xi\Phi_{xx} + \eta\Phi_{yx})\vec{e}_x + 2(\xi\Phi_{xy} + \eta\Phi_{yy})\vec{e}_y =$$
$$= 2\rho(\xi\omega_{\eta\eta} - \eta\omega_{\xi\eta})\vec{e}_x + 2\rho(\eta\omega_{\xi\xi} - \xi\omega_{\xi\eta})\vec{e}_y,$$

$$\frac{1}{4\rho^2}\frac{\sigma^4}{\alpha^4}|\nabla_r\tau^2|^2 = (\xi\omega_{\eta\eta} - \eta\omega_{\xi\eta})^2 + (\eta\omega_{\xi\xi} - \xi\omega_{\xi\eta})^2 = \xi^2\omega_{\eta\eta}^2 - 2\xi\eta\omega_{\eta\eta}\omega_{\xi\eta} + \eta^2\omega_{\xi\eta}^2 + \eta^2\omega_{\xi\xi}^2 -$$
$$-2\xi\eta\omega_{\xi\xi}\omega_{\xi\eta} + \xi^2\omega_{\xi\eta}^2 = \xi^2\omega_{\eta\eta}^2 + \eta^2\omega_{\xi\xi}^2 - 2\xi\eta\omega_{\xi\eta}(\omega_{\xi\xi} + \omega_{\eta\eta}) + \omega_{\xi\eta}^2(\xi^2 + \eta^2),$$

$$\frac{1}{4}|\nabla_r\tau^2|^2 = \frac{\alpha^4}{\sigma^4}\rho^2\left[\xi^2\omega_{\eta\eta}^2 + \eta^2\omega_{\xi\xi}^2 - 2\xi\eta\omega_{\xi\eta}(\omega_{\xi\xi} + \omega_{\eta\eta}) + \omega_{\xi\eta}^2(\xi^2 + \eta^2)\right]. \tag{4.11}$$

Let us calculate $\Delta_r \tau^2$

$$\frac{\sigma^2}{\alpha^2}\Delta_r \tau^2 = \left[2\rho(\xi\omega_{\eta\eta} - \eta\omega_{\xi\eta})\right]_x + \left[2\rho(\eta\omega_{\xi\xi} - \xi\omega_{\xi\eta})\right]_y, \tag{4.12}$$

Let us find each summand in the expression (4.12) separately. For the first summand, we obtain

$$\left[\rho(\xi\omega_{\eta\eta} - \eta\omega_{\xi\eta})\right]_x = \rho_x(\xi\omega_{\eta\eta} - \eta\omega_{\xi\eta}) + \rho(\xi\omega_{\eta\eta} - \eta\omega_{\xi\eta})_x, \tag{4.13}$$

$$\rho_x = -\frac{(\omega_{\xi\xi\xi}\xi_x + \omega_{\xi\xi\eta}\eta_x)\omega_{\eta\eta} + \omega_{\xi\xi}(\omega_{\eta\eta\xi}\xi_x + \omega_{\eta\eta\eta}\eta_x) - 2\omega_{\xi\eta}(\omega_{\xi\eta\xi}\xi_x + \omega_{\xi\eta\eta}\eta_x)}{(\omega_{\xi\xi}\omega_{\eta\eta} - \omega_{\xi\eta}^2)^2} =$$
$$= -\rho^3\left[(\omega_{\xi\xi\xi}\omega_{\eta\eta} - \omega_{\xi\xi\eta}\omega_{\xi\eta})\omega_{\eta\eta} + \omega_{\xi\xi}(\omega_{\eta\eta\xi}\omega_{\eta\eta} - \omega_{\eta\eta\eta}\omega_{\xi\eta}) - 2\omega_{\xi\eta}(\omega_{\xi\eta\xi}\omega_{\eta\eta} - \omega_{\xi\eta\eta}\omega_{\xi\eta})\right]. \tag{4.14}$$

$$(\xi\omega_{\eta\eta} - \eta\omega_{\xi\eta})_x = \xi_x\omega_{\eta\eta} + \xi(\omega_{\eta\eta\xi}\xi_x + \omega_{\eta\eta\eta}\eta_x) - \eta_x\omega_{\xi\eta} - \eta(\omega_{\xi\eta\xi}\xi_x + \omega_{\xi\eta\eta}\eta_x) =$$
$$= \rho\left[\omega_{\eta\eta}^2 + \omega_{\xi\eta}^2 + \xi(\omega_{\eta\eta\xi}\omega_{\eta\eta} - \omega_{\eta\eta\eta}\omega_{\eta\xi}) - \eta(\omega_{\xi\eta\xi}\omega_{\eta\eta} - \omega_{\xi\eta\eta}\omega_{\eta\xi})\right]. \tag{4.15}$$

Substituting (4.14) and (4.15) into (4.13), we obtain

$$\left[\rho\left(\xi\omega_{\eta\eta}-\eta\omega_{\xi\eta}\right)\right]_{x}=\rho^{2}\left[\omega_{\eta\eta}^{2}+\omega_{\xi\eta}^{2}+\xi\left(\omega_{\eta\eta\xi}\omega_{\eta\eta}-\omega_{\eta\eta\eta}\omega_{\eta\xi}\right)-\eta\left(\omega_{\xi\eta\xi}\omega_{\eta\eta}-\omega_{\xi\eta\eta}\omega_{\eta\xi}\right)\right]-$$
$$-\rho^{3}\left[\left(\omega_{\xi\xi\xi}\omega_{\eta\eta}-\omega_{\xi\xi\eta}\omega_{\xi\eta}\right)\omega_{\eta\eta}+\omega_{\xi\xi}\left(\omega_{\eta\eta\xi}\omega_{\eta\eta}-\omega_{\eta\eta\eta}\omega_{\xi\eta}\right)-2\omega_{\xi\eta}\left(\omega_{\xi\eta\xi}\omega_{\eta\eta}-\omega_{\xi\eta\eta}\omega_{\xi\eta}\right)\right]\left(\xi\omega_{\eta\eta}-\eta\omega_{\xi\eta}\right).$$
(4.16)

Similarly, for the second summand from (4.12), we obtain

$$\left[\rho\left(\eta\omega_{\xi\xi}-\xi\omega_{\xi\eta}\right)\right]_{y}=\rho_{y}\left(\eta\omega_{\xi\xi}-\xi\omega_{\xi\eta}\right)+\rho\left(\eta\omega_{\xi\xi}-\xi\omega_{\xi\eta}\right)_{y},$$
(4.17)

$$\rho_{y}=-\frac{\left(\omega_{\xi\xi\xi}\xi_{y}+\omega_{\xi\xi\eta}\eta_{y}\right)\omega_{\eta\eta}+\omega_{\xi\xi}\left(\omega_{\eta\eta\xi}\xi_{y}+\omega_{\eta\eta\eta}\eta_{y}\right)-2\omega_{\xi\eta}\left(\omega_{\xi\eta\xi}\xi_{y}+\omega_{\xi\eta\eta}\eta_{y}\right)}{\left(\omega_{\xi\xi}\omega_{\eta\eta}-\omega_{\xi\eta}^{2}\right)^{2}}=$$
$$=-\rho^{3}\left[\left(\omega_{\xi\xi\eta}\omega_{\xi\xi}-\omega_{\xi\xi\xi}\omega_{\xi\eta}\right)\omega_{\eta\eta}+\omega_{\xi\xi}\left(\omega_{\eta\eta\eta}\omega_{\xi\xi}-\omega_{\eta\eta\xi}\omega_{\xi\eta}\right)-2\omega_{\xi\eta}\left(\omega_{\xi\eta\eta}\omega_{\xi\xi}-\omega_{\xi\eta\xi}\omega_{\xi\eta}\right)\right],$$
(4.18)

$$\left(\eta\omega_{\xi\xi}-\xi\omega_{\xi\eta}\right)_{y}=u_{yy}\omega_{\xi\xi}+\eta\left(\omega_{\xi\xi\xi}u_{xy}+\omega_{\xi\xi\eta}u_{yy}\right)-u_{xy}\omega_{\xi\eta}-\xi\left(\omega_{\xi\eta\xi}u_{xy}+\omega_{\xi\eta\eta}u_{yy}\right)=$$
$$=\rho\left[\omega_{\xi\xi}^{2}+\omega_{\xi\eta}^{2}+\eta\left(\omega_{\xi\xi\eta}\omega_{\xi\xi}-\omega_{\xi\xi\xi}\omega_{\xi\eta}\right)-\xi\left(\omega_{\xi\eta\eta}\omega_{\xi\xi}-\omega_{\xi\eta\xi}\omega_{\xi\eta}\right)\right],$$
(4.19)

Substituting (4.18) and (4.19) into (4.17), we obtain

$$\left[\rho\left(\eta\omega_{\xi\xi}-\xi\omega_{\xi\eta}\right)\right]_{y}=\rho^{2}\left[\omega_{\xi\xi}^{2}+\omega_{\xi\eta}^{2}+\eta\left(\omega_{\xi\xi\eta}\omega_{\xi\xi}-\omega_{\xi\xi\xi}\omega_{\xi\eta}\right)-\xi\left(\omega_{\xi\eta\eta}\omega_{\xi\xi}-\omega_{\xi\eta\xi}\omega_{\xi\eta}\right)\right]-$$
$$-\rho^{3}\left[\left(\omega_{\xi\xi\eta}\omega_{\xi\xi}-\omega_{\xi\xi\xi}\omega_{\xi\eta}\right)\omega_{\eta\eta}+\omega_{\xi\xi}\left(\omega_{\eta\eta\eta}\omega_{\xi\xi}-\omega_{\eta\eta\xi}\omega_{\xi\eta}\right)-2\omega_{\xi\eta}\left(\omega_{\xi\eta\eta}\omega_{\xi\xi}-\omega_{\xi\eta\xi}\omega_{\xi\eta}\right)\right]\left(\eta\omega_{\xi\xi}-\xi\omega_{\xi\eta}\right).$$
(4.20)

As a result, both summand for the expression (4.12) are found. Let us substitute (4.16) and (4.20) into (4.12), we obtain

$$\frac{\sigma^{2}}{2\alpha^{2}}\Delta_{r}\tau^{2}=\left[\rho\left(\xi\omega_{\eta\eta}-\eta\omega_{\xi\eta}\right)\right]_{x}+\left[\rho\left(\eta\omega_{\xi\xi}-\xi\omega_{\xi\eta}\right)\right]_{y}=$$
$$=\rho^{2}\left[\omega_{\eta\eta}^{2}+\omega_{\xi\eta}^{2}+\xi\left(\omega_{\eta\eta\xi}\omega_{\eta\eta}-\omega_{\eta\eta\eta}\omega_{\eta\xi}\right)-\eta\left(\omega_{\xi\eta\xi}\omega_{\eta\eta}-\omega_{\xi\eta\eta}\omega_{\eta\xi}\right)\right]+$$
$$+\rho^{2}\left[\omega_{\xi\xi}^{2}+\omega_{\xi\eta}^{2}+\eta\left(\omega_{\xi\xi\eta}\omega_{\xi\xi}-\omega_{\xi\xi\xi}\omega_{\xi\eta}\right)-\xi\left(\omega_{\xi\eta\eta}\omega_{\xi\xi}-\omega_{\xi\eta\xi}\omega_{\xi\eta}\right)\right]-$$
$$-\rho^{3}\left[\left(\omega_{\xi\xi\xi}\omega_{\eta\eta}-\omega_{\xi\xi\eta}\omega_{\xi\eta}\right)\omega_{\eta\eta}+\omega_{\xi\xi}\left(\omega_{\eta\eta\xi}\omega_{\eta\eta}-\omega_{\eta\eta\eta}\omega_{\xi\eta}\right)-2\omega_{\xi\eta}\left(\omega_{\xi\eta\xi}\omega_{\eta\eta}-\omega_{\xi\eta\eta}\omega_{\xi\eta}\right)\right]\left(\xi\omega_{\eta\eta}-\eta\omega_{\xi\eta}\right)-$$
$$-\rho^{3}\left[\left(\omega_{\xi\xi\eta}\omega_{\xi\xi}-\omega_{\xi\xi\xi}\omega_{\xi\eta}\right)\omega_{\eta\eta}+\omega_{\xi\xi}\left(\omega_{\eta\eta\eta}\omega_{\xi\xi}-\omega_{\eta\eta\xi}\omega_{\xi\eta}\right)-2\omega_{\xi\eta}\left(\omega_{\xi\eta\eta}\omega_{\xi\xi}-\omega_{\xi\eta\xi}\omega_{\xi\eta}\right)\right]\left(\eta\omega_{\xi\xi}-\xi\omega_{\xi\eta}\right),$$

as a result

$$\frac{\sigma^2}{2\alpha^2\rho^2}\Delta_r\tau^2 = \left(\omega_{\eta\eta}^2 + 2\omega_{\xi\eta}^2 + \omega_{\xi\xi}^2\right) + \xi\left[\omega_{\xi\eta\eta}\left(\omega_{\eta\eta} - \omega_{\xi\xi}\right) + \omega_{\xi\eta}\left(\omega_{\xi\xi\eta} - \omega_{\eta\eta\eta}\right)\right] +$$
$$+\eta\left[\omega_{\xi\xi\eta}\left(\omega_{\xi\xi} - \omega_{\eta\eta}\right) + \omega_{\xi\eta}\left(\omega_{\xi\eta\eta} - \omega_{\xi\xi\xi}\right)\right] +$$
$$+\rho\xi\left(\omega_{\eta\eta\eta}\omega_{\xi\xi} + \omega_{\xi\xi\eta}\omega_{\eta\eta} - 2\omega_{\xi\eta}\omega_{\xi\eta\eta}\right)\omega_{\xi\eta}\Delta\omega +$$
$$+\rho\eta\left(\omega_{\xi\xi\xi}\omega_{\eta\eta} + \omega_{\xi\eta\eta}\omega_{\xi\xi} - 2\omega_{\xi\eta}\omega_{\xi\xi\eta}\right)\omega_{\xi\eta}\Delta\omega -$$
$$-\rho\xi\left(\omega_{\xi\xi\xi}\omega_{\eta\eta} + \omega_{\xi\eta\eta}\omega_{\xi\xi} - 2\omega_{\xi\eta}\omega_{\xi\xi\eta}\right)\left(\omega_{\eta\eta}^2 + \omega_{\xi\eta}^2\right) +$$
$$-\rho\eta\left(\omega_{\eta\eta\eta}\omega_{\xi\xi} + \omega_{\xi\xi\eta}\omega_{\eta\eta} - 2\omega_{\xi\eta}\omega_{\xi\eta\eta}\right)\left(\omega_{\xi\xi}^2 + \omega_{\xi\eta}^2\right).$$
(4.21)

Substituting (4.11) and (2.21) into (4.10), we obtain the expression for the quantum potential

$$-\frac{2\beta\sigma^2}{\alpha^3\rho^2}Q = \frac{\sigma^2}{2\alpha^2\rho^2}\Delta_r\tau^2 - \frac{1}{2}\frac{\alpha^2}{\sigma^2}\frac{\sigma^4}{4\alpha^4\rho^2}\left|\nabla_r\tau^2\right|^2 =$$
(4.22)
$$= \omega_{\eta\eta}^2 + 2\omega_{\xi\eta}^2 + \omega_{\xi\xi}^2 +$$
$$+\xi\left[\omega_{\xi\eta\eta}\left(\omega_{\eta\eta} - \omega_{\xi\xi}\right) + \omega_{\xi\eta}\left(\omega_{\xi\xi\eta} - \omega_{\eta\eta\eta}\right)\right] + \eta\left[\omega_{\xi\xi\eta}\left(\omega_{\xi\xi} - \omega_{\eta\eta}\right) + \omega_{\xi\eta}\left(\omega_{\xi\eta\eta} - \omega_{\xi\xi\xi}\right)\right] +$$
$$+\left[\xi\left(\omega_{\eta\eta\eta}\omega_{\xi\xi} + \omega_{\xi\xi\eta}\omega_{\eta\eta} - 2\omega_{\xi\eta}\omega_{\xi\eta\eta}\right) + \frac{\alpha^2}{\sigma^2\rho}\xi\eta + \eta\left(\omega_{\xi\xi\xi}\omega_{\eta\eta} + \omega_{\xi\eta\eta}\omega_{\xi\xi} - 2\omega_{\xi\eta}\omega_{\xi\xi\eta}\right)\right]\rho\omega_{\xi\eta}\Delta\omega -$$
$$-\rho\xi\left(\omega_{\xi\xi\xi}\omega_{\eta\eta} + \omega_{\xi\eta\eta}\omega_{\xi\xi} - 2\omega_{\xi\eta}\omega_{\xi\xi\eta} + \frac{\alpha^2}{2\sigma^2\rho}\xi\right)\left(\omega_{\eta\eta}^2 + \omega_{\xi\eta}^2\right) +$$
$$-\rho\eta\left(\omega_{\eta\eta\eta}\omega_{\xi\xi} + \omega_{\xi\xi\eta}\omega_{\eta\eta} - 2\omega_{\xi\eta}\omega_{\xi\eta\eta} + \frac{\alpha^2}{2\sigma^2\rho}\eta\right)\left(\omega_{\xi\xi}^2 + \omega_{\xi\eta}^2\right).$$

The summands in the expression (4.22) are expressed in terms of the variables $(\xi,\eta)$, and the solution $\tilde{\omega}(\tau,\theta)$ (3.17) is expressed in terms of the variables $(\tau,\theta)$. For the derivatives $\omega_{\xi\xi}$, $\omega_{\xi\eta}$ and $\omega_{\eta\eta}$ the expressions (1.17) were obtained earlier in terms of the derivatives of the solution $\tilde{\omega}_{\tau\tau}$, $\tilde{\omega}_{\theta\theta}$, $\tilde{\omega}_{\tau\theta}$, $\tilde{\omega}_{\tau}$ and $\tilde{\omega}_{\theta}$. In the expression (4.22) in addition to second-order derivatives third-order derivatives $\omega_{\xi\xi\xi}, \omega_{\xi\eta\eta}\omega_{\xi\xi\eta}, \omega_{\eta\eta\eta}$ are also present which must to be expressed in terms of the derivatives of the solution $\tilde{\omega}(\tau,\theta)$. Let us obtain the expressions for the third-order derivatives.

$$\frac{\sigma^3}{\alpha^3}\omega_{\xi\xi\xi} = \tilde{\omega}_{\tau\tau\tau}\cos^3\theta - \tilde{\omega}_{\theta\theta\theta}\frac{\sin^3\theta}{\tau^3} - 3\tilde{\omega}_{\tau\tau\theta}\frac{\cos^2\theta\sin\theta}{\tau} + 3\tilde{\omega}_{\tau\theta\theta}\frac{\cos\theta\sin^2\theta}{\tau^2} +$$
$$+ 3\tilde{\omega}_{\tau\tau}\frac{\cos\theta\sin^2\theta}{\tau} - 6\tilde{\omega}_{\theta\theta}\frac{\sin^2\theta\cos\theta}{\tau^3} + 3\tilde{\omega}_{\tau\theta}\frac{\sin\theta(3\cos^2\theta - 1)}{\tau^2} -$$
(4.23)
$$- 3\tilde{\omega}_{\tau}\frac{\sin^2\theta\cos\theta}{\tau^2} - 2\tilde{\omega}_{\theta}\frac{\sin\theta(4\cos^2\theta - 1)}{\tau^3},$$

$$\frac{\sigma^3}{\alpha^3}\omega_{\eta\eta\eta} = \tilde{\omega}_{\tau\tau\tau}\sin^3\theta + \tilde{\omega}_{\theta\theta\theta}\frac{\cos^3\theta}{\tau^3} + 3\tilde{\omega}_{\tau\tau\theta}\frac{\sin^2\theta\cos\theta}{\tau} + 3\tilde{\omega}_{\tau\theta\theta}\frac{\sin\theta\cos^2\theta}{\tau^2} +$$
$$+ 3\tilde{\omega}_{\tau\tau}\frac{\sin\theta\cos^2\theta}{\tau} - 6\tilde{\omega}_{\theta\theta}\frac{\cos^2\theta\sin\theta}{\tau^3} - 3\tilde{\omega}_{\tau\theta}\frac{\cos\theta(3\sin^2\theta-1)}{\tau^2} - \quad (4.24)$$
$$- 3\tilde{\omega}_{\tau}\frac{\cos^2\theta\sin\theta}{\tau^2} + 2\tilde{\omega}_{\theta}\frac{\cos\theta(4\sin^2\theta-1)}{\tau^3},$$

$$\frac{\sigma^3}{\alpha^3}\omega_{\xi\xi\eta} = \tilde{\omega}_{\tau\tau\tau}\sin\theta\cos^2\theta + \tilde{\omega}_{\theta\theta\theta}\frac{\sin^2\theta\cos\theta}{\tau^3} + \tilde{\omega}_{\tau\tau\theta}\frac{\cos\theta(1-3\sin^2\theta)}{\tau} +$$
$$+ \tilde{\omega}_{\tau\theta\theta}\frac{\sin\theta(1-3\cos^2\theta)}{\tau^2} + \tilde{\omega}_{\tau\tau}\frac{\sin\theta(1-3\cos^2\theta)}{\tau} + 2\tilde{\omega}_{\theta\theta}\frac{\sin\theta(3\cos^2\theta-1)}{\tau^3} + \quad (4.25)$$
$$+ \tilde{\omega}_{\tau\theta}\frac{\cos\theta(9\sin^2\theta-1)}{\tau^2} + \tilde{\omega}_{\tau}\frac{\sin\theta(3\cos^2\theta-1)}{\tau^2} - 2\tilde{\omega}_{\theta}\frac{\cos\theta(4\sin^2\theta-1)}{\tau^3},$$

$$\frac{\sigma^3}{\alpha^3}\omega_{\eta\eta\xi} = \tilde{\omega}_{\tau\tau\tau}\cos\theta\sin^2\theta - \tilde{\omega}_{\theta\theta\theta}\frac{\cos^2\theta\sin\theta}{\tau^3} - \tilde{\omega}_{\tau\tau\theta}\frac{\sin\theta(1-3\cos^2\theta)}{\tau} +$$
$$+ \tilde{\omega}_{\tau\theta\theta}\frac{\cos\theta(1-3\sin^2\theta)}{\tau^2} + \tilde{\omega}_{\tau\tau}\frac{\cos\theta(1-3\sin^2\theta)}{\tau} + 2\tilde{\omega}_{\theta\theta}\frac{\cos\theta(3\sin^2\theta-1)}{\tau^3} - \quad (4.26)$$
$$- \tilde{\omega}_{\tau\theta}\frac{\sin\theta(9\cos^2\theta-1)}{\tau^2} + \tilde{\omega}_{\tau}\frac{\cos\theta(3\sin^2\theta-1)}{\tau^2} + 2\tilde{\omega}_{\theta}\frac{\sin\theta(4\cos^2\theta-1)}{\tau^3}.$$

Substituting (4.23-26), (3.17) into (4.22), we obtain the expression for the quantum potential in terms of the variables $(\tau,\theta)$. Using the expressions (4.22), (4.23-26), (3.17) one can plot the graph of the quantum potential Q and the potential $U$ (1.4) entering the Schrödinger equation. Knowing the potential Q and $U$, one can plot the graph of the classical potential $e\chi$ (1.8).

### 4.3 Special cases of potentials

Let us plot the graphs of the potentials Q, $U$, kinetic energy T and the velocity field $\langle\vec{v}\rangle$ under the condition (3.6). In plotting, we use the algorithm described in Section 4.1 and the expression for the potentials obtained in Section 4.2. Let us consider two limiting cases $\sigma=|\alpha|$ and $\sigma=10|\alpha|$. For each case, we consider two values of the parameter $n=2$ and $n=3$. Let us start with the case $\sigma=|\alpha|$ and $n=2$. Fig. 19 shows the distribution of the velocity vector field $\langle\vec{v}\rangle_2(x,y)$. Fig. 19 shows four flows: two incoming and two outgoing flows. As a result, Fig.19 shows the scattering of two counter flows on the classical potential $e\chi=U+Q=W-T$. The value W corresponds to the total energy and is a constant value. The value T corresponds to the kinetic energy and is shown in Fig.20. Thus, the classical potential coincides, up to a constant W, with $-T$. The distribution graph for the classical potential $e\chi$, as well as the distribution graph for the kinetic energy T (see Fig. 20), has an «angular» shape (rectangular base of the pyramid, see Fig. 20). Such a shape of the potential leads to a corresponding redistribution of the velocity vector field $\langle\vec{v}\rangle$ at scattering on this potential. Fig. 19 shows the rectangular shape as a counter line of the velocity $|\langle\vec{v}\rangle|$. Caused by the collision of the flows, the velocity is close to zero in the central region (see Fig. 19, 20).

The quantum potential Q has a multiplier $\dfrac{|\alpha|}{\beta} = \dfrac{\hbar^2}{2m} \ll 1$ and usually has small contribution to the classical potential $e\chi$. In the case under consideration the value $\sigma = |\alpha| = \dfrac{\hbar}{2m}$ which leads to the proximity of the shapes of the potentials Q and $U$ (see Figs. 21, 22).

Figs. 23-26 show the case $\sigma = 10|\alpha|$ and $n = 2$. Figs. 25, 26 show the distribution for the potentials $U$ and Q, respectively. Due to the increase of $\sigma$, the contribution of the quantum potential is negligible. Indeed, the comparison the graph of the potential $U$ (see Fig. 25) with the graph of the kinetic energy T (see Fig. 24) shows that they are very close up to a sign. The difference between them is the quantum potential Q up to a constant W (see Fig. 26).

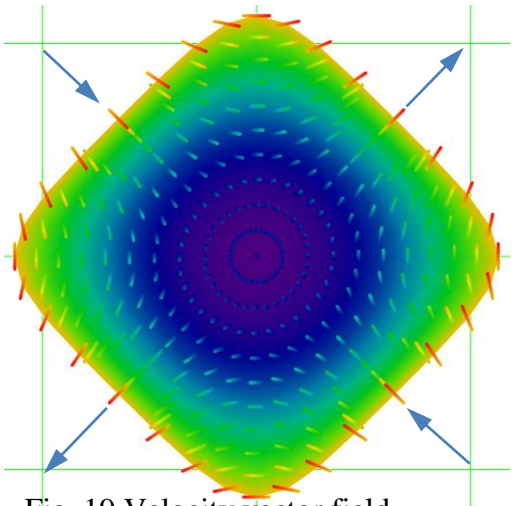

Fig. 19 Velocity vector field $\langle\vec{v}\rangle_2(x, y)$ for $\sigma = |\alpha|$

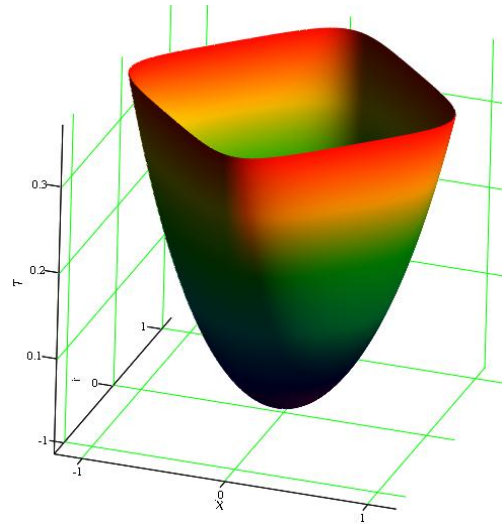

Fig. 20 Kinetic energy $T_2(x, y)$ for $\sigma = |\alpha|$

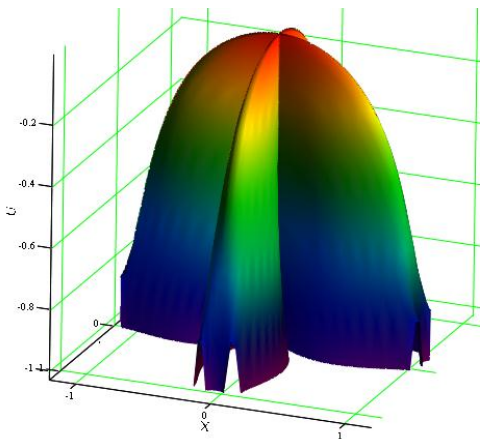

Fig. 21 Potential $U_2(x, y)$ for $\sigma = |\alpha|$

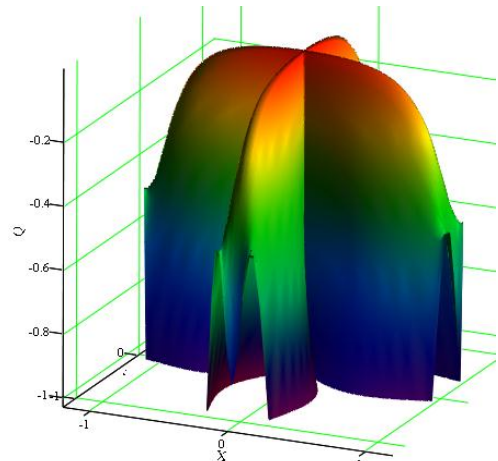

Fig. 22 Potential $-Q_2(x, y)$ for $\sigma = |\alpha|$

Fig. 23 shows the distribution of the velocity vector field for the case $\sigma = 10|\alpha|$ and $n = 2$. Similar to the previous case, there are two incoming and two outgoing flows. Analogously to the previous case, there is a process of scattering of two colliding flows on the potential $e\chi$. However, unlike the previous case, the shape of the potential becomes more «rounded» (the base is the circle of the cone). Such a shape of the potential influences the redistribution of the velocity vector field $\langle \vec{v} \rangle_2 (x, y)$ (see Fig. 23). Comparing the vector field form the previous case (see Fig. 19) with the vector field shown in Fig. 23 it is clear that the counter lines become close to circles.

Let us consider the case $n = 3$. The case $n = 3$ differs significantly from the case $n = 2$. The difference has been noticed in plotting the graphs of the wave functions phases (see Figs. 9, 14). Fig. 9 shows the distribution of the phase $\Phi_2$ of the wave function for $n = 2$. The function $\Phi_2$ is continuously differentiable on the entire mapping region (see Fig. 9).

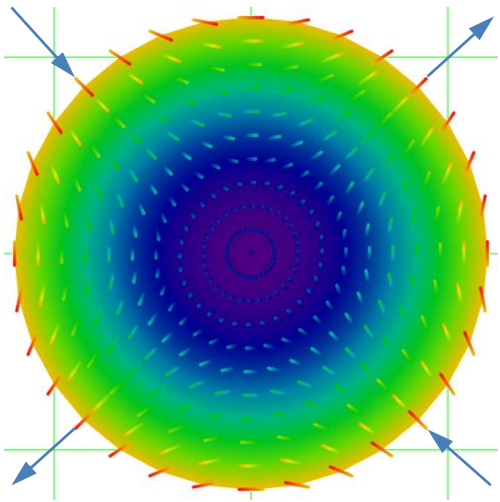

Fig. 23 Velocity vector field $\langle \vec{v} \rangle_2 (x, y)$ for $\sigma = 10|\alpha|$

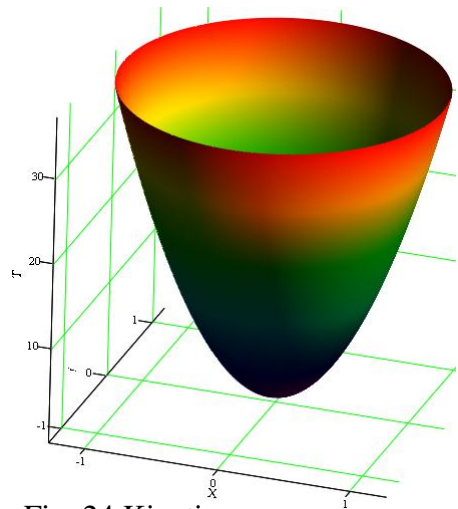

Fig. 24 Kinetic energy $T_2(x, y)$ for $\sigma = 10|\alpha|$

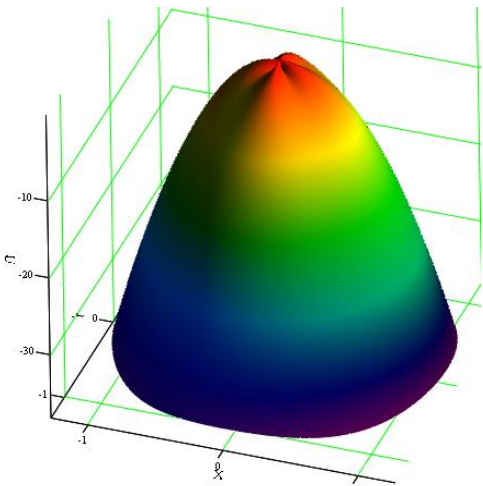

Fig. 25 Potential $U_2(x, y)$ for $\sigma = 10|\alpha|$

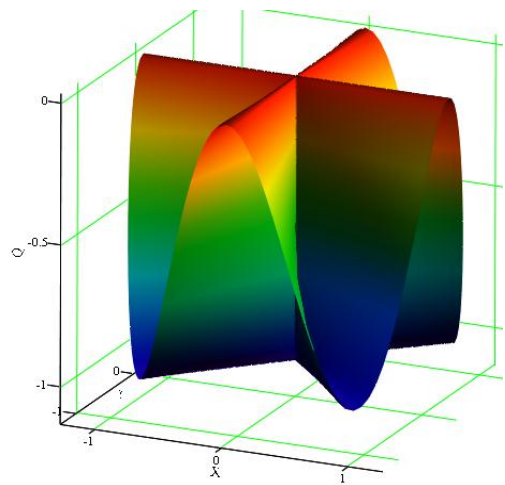

Fig. 26 Potential $-Q_2(x, y)$ for $\sigma = 10|\alpha|$

Fig. 14 shows the distribution of the wave function phase $\Phi_3$ for $n = 3$. Unlike the functions $\Phi_2$, the function $\Phi_3$ is not differentiable for $\theta = 0$. According to the representation (1.10), the velocity vector field $\langle \vec{v} \rangle_3 = -\alpha \nabla \Phi_3$ has a discontinuity for $\theta = 0$. Figs. 27, 31 show the distributions of the vector fields $\langle \vec{v} \rangle_3$ for $\sigma = |\alpha|$ and $\sigma = 10|\alpha|$, respectively. In both cases a discontinuity in the lines of the velocities is seen to the left and to the right of the vertical line corresponding to the angle $\theta = 0$ (see Figs. 27, 31). In the previous case for $n = 2$, as the function $\Phi_2$ is differentiable, there are no such velocity discontinuities observed. Figs. 27, 31 show three flows. There is one incoming and two outgoing flows. One can say that inside the region along the line $\theta = 0$ there is an internal source of the flow caused by the smoothness of the phase $\Phi_3$.

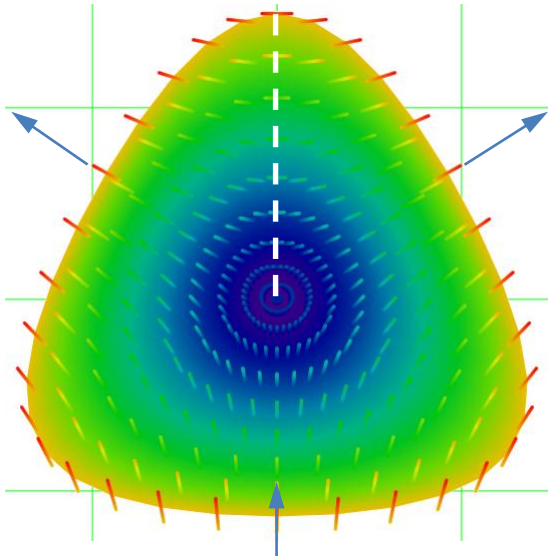

Fig. 27 Velocity vector field $\langle \vec{v} \rangle_3 (x, y)$ for $\sigma = |\alpha|$

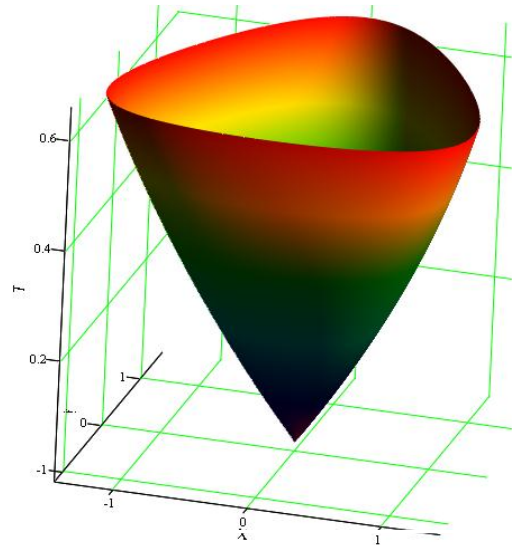

Fig. 28 Kinetic energy $T_3(x, y)$ for $\sigma = |\alpha|$

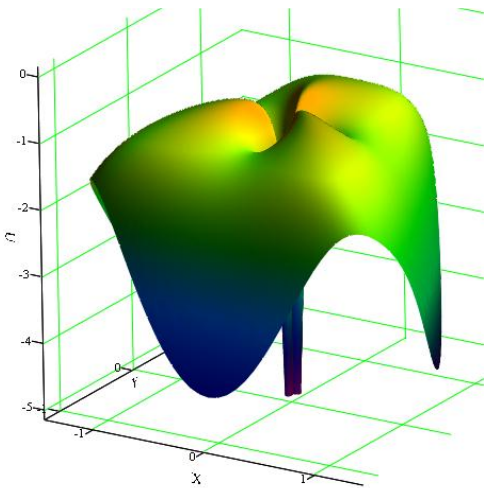

Fig. 29 Potential $U_3(x, y)$ for $\sigma = |\alpha|$

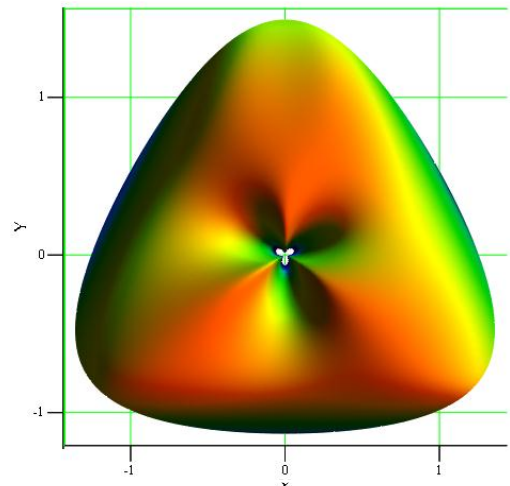

Fig. 30 Potential $-Q_3(x, y)$ for $\sigma = |\alpha|$

As in the previous case ($\sigma = |\alpha|$) an «angular» shape of the distribution of the kinetic energy (triangle base of a pyramid (see Fig. 28)) is observed here. Since the quantum potential for $\sigma = |\alpha|$ has a significant contribution, the graphs $U$ and $Q$ are close in shape. In the case $\sigma = 10|\alpha|$, the influence of the quantum potential is negligible and the shape of the graph of the potential $U$ (see Fig. 33) is close to the shape of the graph of the kinetic energy T (see Fig. 32). The distribution of the vector field modulus has a «rounded» shape (see Fig. 31), which is due to the weak influence of the quantum potential.

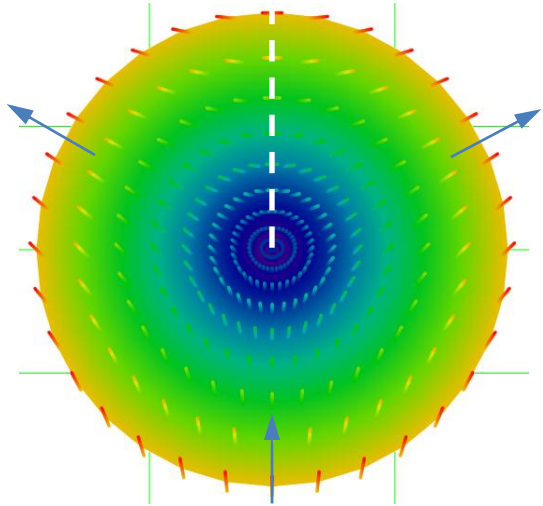

Fig. 31 Velocity vector field $\langle \vec{v} \rangle_3 (x,y)$ for $\sigma = 10|\alpha|$

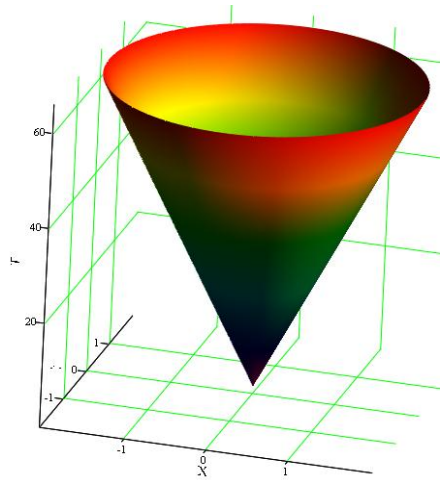

Fig. 32 Kinetic energy $T_3(x,y)$ for $\sigma = 10|\alpha|$

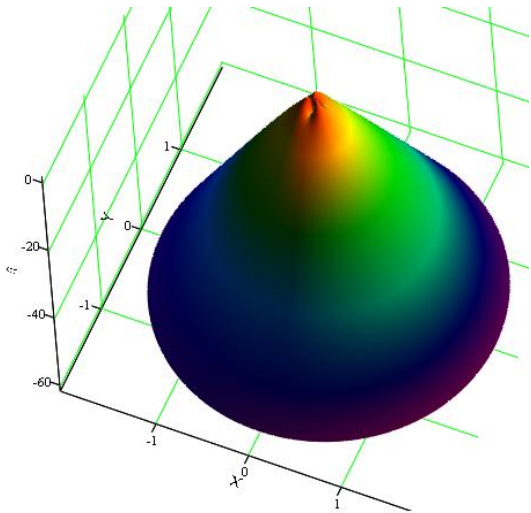

Fig. 33 Potential $\sigma = |\alpha|$ for $\sigma = 10|\alpha|$

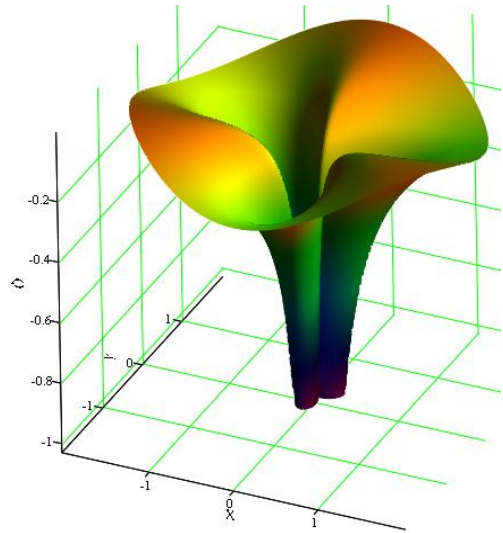

Fig. 34 Potential $-Q_3(x,y)$ for $\sigma = 10|\alpha|$

**Conclusion**

In this paper, we outlined the approach of considering the classical and quantum systems in the frame of a unified mathematical model. As a rule, for transitions between the quantum and classical considerations, certain assumptions are made as well as rejection of quantities of «a higher order of vanishing» and so on. In the approach described in this paper, there were no such «simplifications». The transition between the classical and quantum systems is performed by continuously changing the parameter $\sigma$ entering the Gaussian model distribution (1.9). Although we have considered the distribution (1.9), the approach presented in this paper is applicable to any distribution of a type $f = f(|\langle \vec{v} \rangle|)$.

Despite of the cumbersome formulas in the inverse Legendre transform, the obtained expressions for the probability density, potentials, phases, wave functions, velocity vector fields are exact expressions.